\documentclass[sn-mathphys]{sn-jnl}

\jyear{2021}%

\theoremstyle{thmstyleone}%
\theoremstyle{thmstyletwo}%

\theoremstyle{thmstylethree}%

\usepackage{verbatim} 
\usepackage{tabularx} 

\begin{document}

\title[ALETHEIA: Hunting for Low-mass Dark Matter with Liquid Helium TPCs]{ALETHEIA: Hunting for Low-mass Dark Matter with Liquid Helium TPCs}

\author*[1,2]{\fnm{Junhui} \sur{Liao}}\email{junhui\_liao@brown.edu}

\author[3]{\fnm{Yuanning} \sur{Gao}}\email{yuanning.gao@pku.edu.cn}

\author[2]{\fnm{Zhuo} \sur{Liang}}\email{liangzhuo\_w@163.com}

\author[4]{\fnm{Zebang} \sur{Ouyang}}\email{ ouyangzb1128@163.com }

\author[2]{\fnm{Zhaohua} \sur{Peng}} \email{pzh44@sina.com }

\author[5]{\fnm{Lei} \sur{Zhang}}\email{zlamp@163.com}

\author[2]{\fnm{Lifeng} \sur{Zhang}}\email{zlf20042008@126.com}

\author[2]{\fnm{Jian} \sur{Zheng}}\email{13522656935@139.com}

\author[2]{\fnm{Jiangfeng} \sur{Zhou}}\email{853051644@qq.com}

\affil*[1]{\orgdiv{Department of Physics}, \orgname{Brown University}, \orgaddress{\street{Hope St. 182}, \city{Providence}, \postcode{02912}, \state{Rhode Island}, \country{USA}}}

\affil[2]{\orgdiv{Division of Nuclear Physics}, \orgname{China Institute of Atomic Energy}, \orgaddress{\street{Sanqiang Rd. 1}, \city{Fangshan district}, \postcode{102413}, \state{Beijing}, \country{China}}}

\affil[3]{\orgdiv{School of Physics}, \orgname{Peking University}, \orgaddress{\street{ChengFu Rd. 209}, \city{Haidian district}, \postcode{100084}, \state{Beijing}, \country{China}}}

\affil[4]{\orgdiv{School of Nuclear Technology}, \orgname{University of South China}, \orgaddress{\street{ChangSheng West Rd. 28}, \city{Hengyang}, \postcode{421009}, \state{Hunan}, \country{China}}}

\affil[5]{\orgdiv{Division of Nuclear Synthesis Technology}, \orgname{China Institute of Atomic Energy}, \orgaddress{\street{Sanqiang Rd. 1}, \city{Fangshan district}, \postcode{102413}, \state{Beijing}, \country{China}}}

\abstract{Dark Matter (DM) is one of the most critical questions to be understood and answered in fundamental physics today. Observations with varied astronomical and cosmological technologies strongly indicated that DM exists in the Universe, the Milky Way, and the Solar System. Nevertheless, understanding DM under the language of elementary physics is still in progress. DM direct detection tests the interactive cross-section between galactic DM particles and an underground detector's nucleons. Although Weakly Interactive Massive Particles (WIMPs) are the most discussed DM candidates, the null-WIMPs conclusion has been consistently addressed by the most convincing experiments in the field. Relatively, the low-mass WIMPs region ($\sim$ 10 MeV/c$^2$ - 10 GeV/c$^2$) has not been fully exploited compared to high-mass WIMPs ($\sim$ 10 GeV/c$^2$ - 10 TeV/c$^2$). The ALETHEIA (A Liquid hElium Time projection cHambEr In dArk matter) experiment aims to hunt for low-mass WIMPs with liquid helium-filled TPCs (Time Projection Chambers). In this paper, we go through the physics motivation of the project, the detector's design, the R\&D plan, and the progress we have made since the project has been launched in the summer of 2020.}

\keywords{Dark Matter, low-mass WIMPs, Liquid Helium, TPCs}

\maketitle


\section{Introduction}\label{sec1}
\subsection{The existence of dark matter and its detection}\label{sec1sub1}

Variant astronomical evidence, such as cluster and galaxy rotation curves~\cite{Zwicky33, Rubin70}, lensing studies and spectacular observations of galaxy cluster collisions~\cite{Refreiger03, Clowe06, Fields08}, and Cosmic Microwave Background (CMB) measurements~\cite{Planck2018ResultsOne}, all point to the existence of Cold Dark Matter (CDM) particles. Cosmological simulations based on the CDM model have been remarkably successful at predicting the structures we see in the Universe~\cite{Davis1985, Navarro96}. Alternative explanations proposing a modification of general relativity can explain some limited astronomical observations but have not been able to explain this large body of evidence across all length scales~\cite{feng2014planning}. 
Recent results from Gaia~\cite{GaiaExp} show consistency with many previous experiments and therefore much more securely pinned down than ever on: (a) DM dominates the mass of the Milky Way Galaxy~\cite{PostiHelmi19}, and (b) the local DM mass density in the Solar System is around 0.3 GeV/c$^2$ $\cdot$ cm$^{-3}$~\cite{HagenHelmi18}.

Weakly Interactive Massive Particles (WIMPs) are one of the most prominent DM candidates today, which were first proposed in the 1980s~\cite{SteigmanTurner85}. WIMPs are a hypothesized class of DM particles that would freeze out of thermal equilibrium in the early Universe and result in a relic density matching today's observation. WIMPs are often called a ``miracle.'' With "miracle," it means that given the density of WIMPs inferred by cosmological and astronomical measurements, with a possible mass in the range of $\sim$ 10 - 100 GeV/c$^2$~\cite{Feng10}.

There are several viable strategies to detect DM. First, indirect detection experiments aim to observe high-energy particles resulting from the self-annihilation of DM. Second, collider experiments look for the production of DM particles in high-energy collisions. Finally, direct detection experiments try to observe the rare scatters of DM on the very low background detectors that operate in deep underground laboratories. 

Based on physical motivations, DM direct detection experiments can be roughly classified into four categories: low-mass DM experiments ($\sim$ 10 MeV/c$^2$ - 10 GeV/c$^2$), high-mass DM experiments ($\sim$ 10 GeV/c$^2$ - 10 TeV/c$^2$), annual modulation, and directional detection. Table~\ref{tabDMExpCat} categorizes (most if not all) currently active DM direct detection experiments in the field~\footnote{The classification is not very strict in the sense that some experiments can have more than one categories in the table~\ref{tabDMExpCat}, or even beyond the scope of DM search. Taking LZ as an example, LZ is a high-mass DM experiment, but its physics searches are extensive, including but are not limited to: annual modulation, low-mass WIMPs search via S2O (S2 Only) analysis, Axion-like particles, two neutrinos double beta decay, and neutrinoless double beta decay, etc. Other leading DM experiments have a similar feature.}.

\begin{table}[h]
\begin{minipage}{328pt}
\caption{Categorization of DM direct detection experiments with physics motivation}\label{tabDMExpCat}%
\begin{tabular}{@{}llll@{}}
\toprule
Low-mass & High-mass  & Annual modulation  & Directional \\
\midrule
      ALETHEIA, CEDX,				&ArDM, DarkSide, 		&ANAIS,				& DRIFT, \\
      CRESST, DAMIC, Edelweiss,		&DEAP, LZ, PandaX, 	&COSINE-100, 		& NEWAGE, \\
      SENSI, SuperCDMS.				&PICO, XENON.	 	&DAMA/LIBRA.			& NEWS. \\
\botrule
\end{tabular}
\end{minipage}
\end{table}

\subsection{Low-mass dark matter}\label{sec1sub2}

Independent of the SUSY (SuperSymmetry) models, which predicate candidate particles having the similar features of WIMPs~\cite{Jungman96}, there are four categories of well-motivated scenarios that favor MeV/c$^2$ - GeVc/c$^2$ dark matter.

(a) the ``WIMPs miracle'' model motivates $\sim$ 10 MeV/c$^2$ - 100 TeV/c$^2$ WIMPs~\cite{Feng10, FengKumar08}: a stable particle with such mass annihilating each other with a cross-section of $\sigma v \sim 2 \times 10^{-26} \text{cm}^3/$s in the early university would result in consistent dark matter density as measured~\cite{Steigman12}. DM mass greater than 240 (340) TeV/c$^2$  would violate the requirement of partial wave unitarity assuming DM is a Dirac (Majorana) fermion~\cite{Griest90}, while the Big Bang nucleosynthesis would ruin if an annihilating relic with a mass lighter than $\sim$ 1 - 10 MeV/c$^2$~\cite{Bohm13}.

(b) Light DM annihilates to quarks might have been suppressed during the epoch of recombination; instead of coupling to quarks, light DM would possibly couple to leptons, in particular, electrons. In direct detection, DM could scatter with electrons to generate individual electrons, individual photons, individual ions, and heat/phonons~\cite{Essig12}. 

(c) The asymmetry in the dark sector might be related to the baryon asymmetry (matter and anti-matter), resulting in zero net baryon number of the universe ~\cite{Cohen93, Zurek14}. The mass of DM would be $\sim r \times (4-5)$ GeV/c$^2$, where $r$ is a factor that maintains equilibrium between the dark and the Standard Model (SM) sectors in the early universe. If $r = 1$, the DM mass would be $\sim (4-5)$ GeV/c$^2$. 

(d) Strongly Interacting Massive Particle (SIMP) models propose dark matter as a meson- or baryon-like bound-state of hidden sector particles, with a mass near the QCD scale, $\sim$ 100 MeV/c$^2$~\cite{StrasslerZurek07, Hochberg14, Hochberg15, Kuflik16, Kuflik17}. Considering constraints from the CMB, dark matter masses might be in the range of $\sim$ 5 - 200 MeV/c$^2$. 

\subsection{The interactions between low-mass WIMPs and matter}\label{sec1sub3}

Since WIMPs are supposed to be neutral particles, so the mediators for WIMPs interacting with matter are limited to either the Z boson or the Higgs boson, according to the SM (Standard Model). Depending on the particle characters of WIMPs and the coupling between WIMPs and the mediators, the interactions between WIMPs and matter could be hugely different, as discussed in reference~\cite{Escudero2016}. However, as a general statement,  compared to the high-mass region, the low-mass WIMPs have a more expansive parameter space that remains to be explored. 

In the paper, we limit our discussion to the SI (Spin Independent) framework, which is the dominant model in the DM direct detection community; specifically, we assume WIMPs are fermions, the mediators involved in the interaction are either the Z-boson or the Higgs-boson. Under the SI model, WIMPs would scatter off a DM detector’s nuclei coherently. The scattered nuclei, therefore, receive recoil energy from the WIMPs and register as an NR (Nuclear Recoil) event in the detector. That being said, WIMPs events are NR-like signals. As a comparison, backgrounds events are ER (Electron Recoil) like, which usually result from gammas or electrons interacting with the detector. 

However, ER could also be DM signals under such alternative scenarios as Boosted dark matter, Exothermic dark matter, and Bosonic dark matter~\cite{Leane2022}, as well as DM - electrons direct coupling as mentioned above. In general, the fewer the NR and ER events, the easier to figure out the events' origin: signals or backgrounds. ``As Few events As Possible (AFAP)'' is critical because nobody can currently guarantee what kind of DM signal it is, ER? or NR? So an ideal DM detector should show the features of (a) having the least intrinsic NR and ER background events and (b) being capable of discriminating ER from NR events. Fortunately, the ALETHEIA detector would have these two advantages and beyond. For more details, please refer to section~\ref{secProjectSubWhyLHeTPC}.

\subsection{Experimental progress}\label{sec1sub4}

Experimentally, in the high-mass WIMPs region, the current lowest limits for WIMPs-nucleon interaction with the SI framework are down to $\sim$ 10$^{-48}$ cm$^2$ for $\sim$ 30 GeV/c$^2$ WIMPs~\cite{LZ2022, PandaX2021, Xenon1TPaper18}, which are roughly eight and two orders lower than the interaction hypothetically mediated through the Z boson ($\sim$ 10$^{-39}$ cm$^2$) and the Higgs boson ($\sim$ 10$^{-46}$ cm$^2$), respectively. As a comparison, in the low-mass region, however, there exists broader open parameters space to be exploited; In particular, for WIMPs mass less than 2 GeV/c$^2$, the current limits roughly are $\sim$ 10$^{-38}$ cm$^2$, which are one and six orders higher than the interaction mediated by the Z and Higgs boson, respectively.

Low-mass WIMPs should still be investigated regardless of whether WIMPs could be discovered at the scale of $\sim$ 100 GeV/c$^2$. Because (a) it has been motivated by several different mechanisms, as mentioned in subsection~\ref{sec1sub2}; (b) considering there exist tens of elementary particles in the SM, it is reasonable to hypothesize that existing more than one dark matter particle, which would naturally have different masses. So, even if WIMPs were discovered in the high-mass region, physicists should also check the low-mass region to see if there is low-mass dark matter and vice versa.

However, low-mass WIMPs are out of the scope that leading high-mass DM experiments can effectively reach with ``classical'' analysis such as ``S1/S2''~\footnote{``S1'' refers to the prompt scintillation when incident particles hit the TPC bulk; ``S2'' is the electroluminescence generated at the helium gas layer, which originally comes from the ionized electrons being separated from their neighbor ions by external field then drift to the gas layer. The ratio of S1 and S2 signals' amplitude, or S1/S2, is proved to be an efficient way to discriminate ER from NR events in liquid noble gas experiments. Essentially, the discrimination originates from different charge densities of two types of recoils. Please refer to subsection~\ref{sec4QsRs-sub2S1S2} for more details.} and PSD (Pulse Shape Discrimination). For LXe TPCs, this is because the detector's material (xenon) is quite heavy, and recoil energy is inversely proportional to the atomic number of the detector material. Although argon is not very heavy, the PSD analysis requires a significant number of photoelectrons for LAr experiments, limiting their ROI  (Research Of Interest) to reaching the low-mass region.

\subsection{ALETHEIA and other LHe DM projects}\label{sec1sub5}

The ALETHEIA project aims to hunt for low-mass WIMPs, which was inspired by many respected peer experiments in the community. Although there already exist quite a few low-mass DM experiments~\cite{SuperCDMS2014, SuperCDMS2016, DAMIC2016, CRESST2014, CRESST2017, EDELWEISS2017, CDEX2018}, we believe ALETHEIA is a competitive project in the race of low-mass DM search thanks to the extremely low backgrounds it could potentially achieve. In this paper, without special notice, helium refers to $^4$He.

The most distinctive feature of ALETHEIA is that it could achieve an IBF (Instrumental Background Free) search, which means in the project's ROI, a small number of background events (for instance, $<$ 0.1 events) are expected, and after a series of events selections, zero background events are survived in the ROI region. The instrumental backgrounds include radioactive particles due to the materials in the detector system (including dust) and the particles generated by cosmological muons hitting the rocks near the detector or the detector itself. The DarkSide-50 (DS-50) experiment already demonstrated that an IBF search is viable~\cite{DarkSide502018}. DS-50 relies heavily on LAr's extremely powerful PSD to achieve IBF, among other efforts. For ALETHEIA, some unique advantages are: (a) the lowest intrinsic backgrounds contributed from the detector's bulk material, LHe, and (b) the potentially strong capability of ER/NR discrimination with either the S1/S2, or PSD, or both analyses.

ALETHEIA is not the only project to utilize LHe to hunt for DM. There exist a couple of other projects that aim to implement Superfluid Helium(SHe) to do DM searches in recent years, as shown in references~\cite{Hertel19, SpiceHeRald21}, and~\cite{Maris17}. Helium gas becomes liquid as long as being cooled to 4.5 K; this is the ``general'' LHe. Keep cooling to 2.17 K or below, and LHe reaches the superfluid status. SHe has many different features than the general LHe. One of them is heat conductivity. As a good heat conductor, SHe can detect ``quasiparticles'' (phonons and rotons) left by incident particles. Thanks to the low energy of quasiparticle production, 0.62 meV, a SHe detector can, in principle, detect $\sim$ MeV/c$^2$ DM. However, the SHe detector must work at an extremely low temperature as 100 mK or below~\cite{Hertel19}\footnote{Actually, the temperature for other similar cryogenic detectors is $\sim$10 mK. For instance, the temperature of the CUORE and SuperCDMS detectors are, 10 mK~\cite{CUORE2022} and 15 mK~\cite{SuperCDMS2016}, respectively}. It is more challenging to build a SHe detector than a general LHe one in terms of cryogenic engineering.  

This paper will focus on the ALETHEIA project for the experimental detection and verification of the WIMPs hypothesis. We organize this paper as the following: We first introduce the ALETHEIA project in more detail in section~\ref{secProject}, then we address the technical design for the ALETHEIA detector in section~\ref{secDesignALETHEIA}, we introduce the LHe TPC's R\&D plan in section~\ref{sec4QsRs}, we deliver the progress of the project in section~\ref{secProgressALETHEIA}, we summarize the project in section~\ref{secSummaryALETHEIA}. 


\section{The ALETHEIA project}\label{secProject}

In this section, we first introduce the ALETHEIA project in details in~\ref{secProjectSubROI}. We then explain why an LHe TPC is suitable for a low-mass DM hunting in~\ref{secProjectSubWhyLHeTPC}. Next, we bring up the project review that happened in Oct 2019 in~\ref{secProjectSubProjectReview}. We also discuss the main technical challenges of an LHe TPC we already knew and how the challenges could affect the detector's performance in~\ref{secProjectSubChallenges}. We finally touch what will not be covered in the paper in~\ref{secProjectSubNotCovered}.

\subsection{The ALETHEIA's ROI}\label{secProjectSubROI}

\begin{figure}[!t]	 
	\centering
    \includegraphics[width=1.0\textwidth]{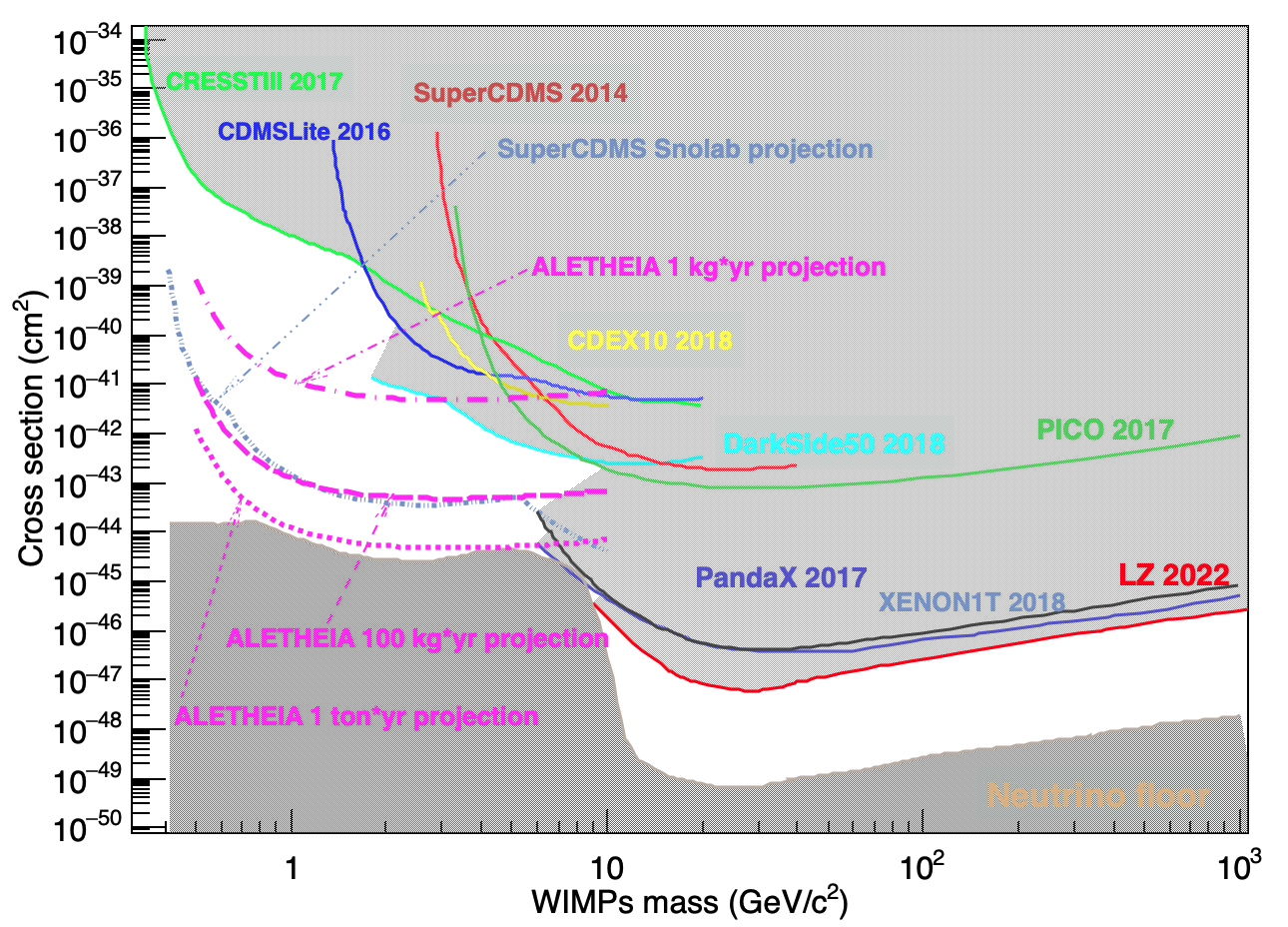}
	\caption{The figure shows the parameter space for the spin-independent WIMPs-nucleon cross-section. The dark green region represents the space that has been excluded by leading direct detection experiments at a 90\% confidence level. The area with light brown color corresponds to the space of the ``neutrino floor'' where neutrinos can generate WIMPs-like events in DM detectors. The projected sensitivities of the ALETHEIA with the exposure of 1 kg*yr, 100 kg*yr, and 1 ton*yr are shown. Zero backgrounds and 50\% detection efficiency were assumed for the projections. Please refer to the main text for more information. The upper limits or projected limits of other experiments are also shown. The ``CRESSTIII-2017'' data cited from~\cite{CRESST2017}, ``CDMSLite-2016'' from~\cite{CDMSLite2016}, ``CDEX10 2018'' from~\cite{CDEX2018}, ``DarkSide50 2018'' from~\cite{DarkSide502018}, ``LZ 2022'' from ~\cite{LZ2022}, ``PandaX 2021'' from~\cite{PandaX2021}, ``PICO 2017'' from~\cite{PICO2017}, ``Neutrino floor'' from~\cite{Billard14}., ``SuperCDMS 2014'' from~\cite{SuperCDMS2014}, ``SuperCDMS Snolab projection'' from~\cite{SuperCDMS2016}, ``XENON1T 2018'' from~\cite{Xenon1TPaper18}.}\label{projectedSensitivities} 
\end{figure}

The ALETHEIA's ROI is low-mass DM. With the SI model and the S1/S2 analysis, the ROI is $\sim$ 500 MeV/c$^2$ - 10 GeV/c$^2$ WIMPs~\footnote{For 500 MeV/c$^2$ WIMPs; here, we assume the velocity WIMPs hitting the detector is 2 $\times$ 10$^{-3}$ c (c is the speed of light), so the kinetic energy is 1 keV, the maximum nuclear recoil energy is 0.4 keV$_{nr}$, the electron equivalent recoil energy is roughly 40 eV$_{ee}$ assuming the QF (Quenching Factor) is 0.1 at 0.4 keV$_{nr}$ (extrapolate from reference~\cite{Santos08}), which is equivalent to the average ionization energy of LHe, 42.3 eV, as shown in table~\ref{tabLHeProperties}. The WIMPs velocity follows an ``irregular'' distribution according to Gaia data~\cite{Necib19}. Choosing 2 $\times$ 10$^{-3}$ c will have a different probability value than 1 $\times$ 10$^{-3}$ c. The probability can eventually be integrated into the detector's detection efficiency.}; accordingly, projected sensitivities of the ALETHEIA with the exposure of 1 kg*yr, 100 kg*yr, and 1 ton*yr are shown in Fig~\ref{projectedSensitivities}. The projected sensitivities are calculated on two assumptions: (a) there is no background in the interesting recoil energy range (after a series of cuts being applied), and (b) the detection efficiency is 50\% for the whole interesting energy interval. We made the assumption (a) because we believe an extremely low background or even IBF search is achievable, as mentioned in section~\ref{sec1sub5} above. Moreover, as will be discussed in section~\ref{secProjectSubChallengesSubsubSlowMobility}, for a 1.5-m, 0.33-ton LHe TPC, a 1 ton*yr exposure would only have 11 ER and 0.5 NR background events, respectively; considering 99.5\% ER rejection and 50\% NR detection efficiency (the same as LZ~\cite{LZTDR17}), for a 1 ton*yr exposure, the background events would be (11 * 0.5\% + 0.5 * 50\%)  = 0.3 events. For 1 kg*yr and 100 kg*yr exposures, the background events are supposed to be smaller than 0.3 due to less radioactive detector materials will be used. We, therefore, made the zero-background assumption for the projections. The detection efficiency is often recoil-energy dependent. Due to the lack of reliable experimental data, we made a flat efficiency of 50\% for the ROI energy interval as a preliminary input to calculate the projections. We will update the projections after we have data on backgrounds and detection efficiency.

In Fig~\ref{projectedSensitivities}, we did not show the projected sensitivities with other paradigms and analyzing methods. For instance, WIMPs might be absorbed by the target nucleus instead of being scattered, as suggested in reference~\cite{Dror20}. In the scenario, the recoil energy is $m_\chi^2/2M$, where $m_\chi$ and M are the mass of DM and target nucleus, respectively. For a 50 MeV/c$^2$ WIMPs, the recoil energy is ``boosted'' to around 300 keV$_{nr}$ for an LHe detector. This model significantly expands the ALETHEIA's sensitive WIMPs mass region, all the way down to $\sim$2 MeV/c$^2$. As to the analysis methods, besides the ``classical'' S1/S2 and PSD, the S2 only (S2O) analysis can also extend the sensitive mass of a detector downwards one to two orders according to DarkSide-50 and XENON results~\cite{DarkSide502018S2O, XENON1TS2O19}. Similarly, applying the S2O for ALETHEIA can, in principle, extend its sensitive DM mass downwards. However, the S2O analysis depends strongly on data; without data in hand, it is not easy to project the sensitivity at the current stage. We will address this in more detailed below~\ref{sec4QsRs-sub2S2OSubS2ODMSearches}.
In the following sections, without special notice, we will mainly stick to the SI model and the S1/S2 and PSD analyses.

For any low-mass WIMPs experiment with an ambition of touching down on the $^8$B ``neutrino floor''~\cite{Billard14} or ``neutrino fog''~\cite{OHare21, NextGenerationLXeObservatory22} cross-section, $^8$B solar neutrinos register an inevitable background. The $^8$B events cannot be discriminated from low-mass WIMPs signals. Nevertheless, according to reference~\cite{Bahcall04, Billard14, SNO02}, the measured $^8$B events are well consistent with the theoretical prediction. The uncertainty of the events is $\sim$16\%, as reported in reference~\cite{SNO02}. As a result, the well-estimated $^8$B events can be considered a known background. Therefore it is still possible to extrapolate WIMPs signal from underground experimental data containing the $^8$B events using the Profile Likelihood Ratio (PLR) analysis~\cite{Cowan2011}~\footnote{Compared to the ``cut and count'' method, which relies only on the number of signal or background events, the PLR also integrates other patterns, such as the detector's response to ER and NR events and uncertainties, to constrain statistical fluctuations, therefore, improve the sensitivity of data: a discovery or upper limit.}. Under the context, the fewer the instrumental backgrounds and uncertainties, the smaller the WIMPs nucleon cross-section can be reached. The (to be demonstrated) IBF feature of ALETHEIA fits perfectly on the searches that $^8$B neutrinos matter.

\subsection{Why LHe TPC?}\label{secProjectSubWhyLHeTPC}

An LHe TPC has several unique advantages, enabling the ALETHEIA detector to achieve an IBF (or extremely low backgrounds) search.

(a) High recoil energy. Helium is the lightest noble element and the second light element. Therefore, the same kinetic energy of incident WIMPs would result in greater recoil energy than other elements except for hydrogen. Although hydrogen is even lighter, hydrogen is not a good detector material due to a few drawbacks such as explosive, chemically active, etc. 

(b) High QF (Quenching Factor). For 16 keV nuclear recoil energy, the measured QF of LHe is $\sim$ 65\%~\cite{Santos08}; while LAr is $\sim$ 24\%~\cite{QFArgon14}, which is a factor of 3 smaller. Another measured QF of helium at 1.5 keV recoil energy is up to 22\%~\cite{Santos08}. As a comparison, the measured QF of hydrogen at 100 keV is only 2\%~\cite{Reichhart12}, and the estimated QF at 1.5 keV nuclear recoils would be much lower than 2\%. As a result, the QF of hydrogen is guaranteed to be at least one order smaller than helium at $\sim$1 keV$_{\text{nr}}$. Thus, for the same incident kinetic energy, although hydrogen has 4-fold greater recoil energy than helium, by considering their quenching factors, the electron equivalent recoil energy of hydrogen is at least a factor of 2.5 smaller than helium. 

One should also be aware that in the QF-measurement paper~\cite{Santos08}, only ionization is registered with a Micromegas detector, and excitation-induced scintillation is ignored, which is different from our detector. In our case, an LHe TPC can collect scintillation produced both by excitation and ionization. Consequently, the scintillation yield in an LHe TPC should be greater than the quenched energy alone since the process of excitation and de-excitation can also contribute scintillation.

(c) Radioactivity free. Helium has no radioactive isotopes; both $^3$He and $^4$He are stable elements. However, LAr and LXe have radioactive isotopes, and most (if not all) of these isotopes are very difficult to eliminate.

(d) Easy to purify. At 4 K temperature, $^3$He is the only solvable material in LHe~\footnote{Hydrogen can also solve in LHe, but the solubility is as low as 10$^{-14}$~\cite{JEWELL1979682} at 1 K. Hydrogen is not a radioactive element so it will not contribute background events.} while it is very rare, the ratio of $^4$He and $^3$He in Nature is $10^7$:1; any other impurities would exist in the solid-state and, therefore, easy to be purified with getters and cold-traps. LAr and LXe have solvable impurities and are not easy to get purified, although the purification level is continuously improved~\cite{XENONnTERAnalysis2022, XENONnT2022IDMTalk}. Actually, one of the most important advantages of liquid noble gas TPCs is that they can purify their bulk material online and keep it extremely clean, while solid detectors can not.

(e) Possible strong ER/NR discrimination. TPC technology is well-developed in LXe and LAr experiments and helped these projects to take a leading role in high-mass DM search in the past decade, one after another. These experiments demonstrated that the S1/S2 and/or PSD analysis could be implemented to achieve efficient ER/NR discrimination. As a TPC filled with another noble gas, helium; hopefully, ALETHEIA could mimic the success of peer experiments and help to explore the parameters space other experiments might be difficult to touch.

(f) Scalability. LXe and LAr experiments demonstrated that liquefied noble gas-filled TPCs could scale from $\sim$ kg up to $\sim$10 ton, even multi-10-ton. Therefore, we hope that an LHe TPC can also scale up to be a 1-ton and multi-ton detector, with which ALETHEIA could fully touch down the B-8 neutrino floor (fog) or even below. 

And (g) Helium procurement. Helium is significantly cheaper than xenon. The price of helium is $\sim$ 1/7 of xenon (Though the prices of helium and xenon both fluctuate.). Every year, helium consumption is $\sim$ 20 k ton globally~\cite{heliumPriceUCSB}. One can buy 1-ton LHe without affecting the market price since it is only 0.005\% of the total consumption.\\

Although the reference~\cite{GuoMckinsey13} proposed the concept of an LHe TPC, no such detector has ever been built worldwide. As a creative technology, an LHe TPC naturally has some risks. To better understand the technical risks and challenges, we invited a few leading physicists in the field of LHe, TPC, and DM to review the ALETHEIA project in Oct 2019~\cite{DMWS2019PKU}. The details of the review will be introduced in section~\ref{secProjectSubProjectReview}.

\subsection{The project review}\label{secProjectSubProjectReview}

In Oct 2019, we organized a DM workshop at Peking University in Beijing, China~\cite{DMWS2019PKU}. The corresponding author of the paper, Dr. Liao, presented the project's concept during the workshop~\footnote{The then name of the project is ALHET which stands for A Liquid HElium Time projection chamber.}~\cite{LiaoPresentationDMWS2019PKU} and provided a $\sim$20-page documents (main text)~\cite{ALETHEIA_2019WorkPlan} to address the detailed R\&D plan for the project. A panel composed of leading physicists~\footnote{The review panel members are: Prof. Rick Gaitskell at Brown University, Prof. Dan Hooper at Fermilab and the University of Chicago, Dr. Takeyaso Ito at Los Alamos National Laboratory, Prof. Jia Liu at Peking University, Prof. Dan McKinsey at UC Berkeley and Lawrence Berkeley National Laboratory, and Prof. George Seidel at Brown University.} in DM direct detection, liquid helium, and DM theory reviewed the project. According to the reviewing panel~\cite{ALETHEIA_2019Review}, the LHe TPC technology is very competitive, ``\textit{It is possible that liquid helium (TPC) could enable especially low backgrounds because of its powerful combination of intrinsically low radioactivity, ease of purification, and charge/light discrimination capability.}''~\cite{ALETHEIA_2019Review}. 

Meanwhile, the panel addressed three categories of challenges.

(i), Low-energy recoils corresponding small signals in an LHe detector. The ALETHEIA's ROI is 100s MeV/c$^2$ - 10 GeV/c$^2$. No matter which analyzing method(s) will be implemented at the end, the detector shares the same challenges: it must be capable of detecting single photoelectron size S1 and the single electrons generated by primary ionization for S2. However, preliminary calculations and estimations show both requirements are viable, as detailed in subsection~\ref{sec4QsRs-sub2S1S2subALETHEIA}. A complete simulation on liquid helium excitation, ionization, photons propagation, etc., should be supplied in the future.

(ii), Electrons induced bubbles. In LHe, as an electron's kinetic energy decreases to approximately 1 eV, the electron would become a bubble with a radius of $\sim$ 19 \r{A}~\cite{Maris08} \footnote{Similarly, an energized ion would produce a snowball with a radius of $\sim$ 6-7\r{A}. Since a TPC mainly focuses on the mobility of electrons, we will not discuss snowballs here.}. The bubble's mobility is slow and electric field dependent, 2 $\times 10 ^{-2}$ cm$^2$ / (V $\cdot$ s) at 4.0 K~\cite{Schwarz72}. In a field of $10^4$ V/cm, the velocity of electrons is 2 m/s. Fortunately, such a slow velocity will not bother the operation of an LHe TPC, thanks to the extremely low intrinsic backgrounds. For details, please refer to subsection~\ref{secProjectSubChallengesSubsubSlowMobility}.

(iii) HV (High Voltage) related concerns. As mentioned, to make electrons have a reasonable mobile velocity of 2 m/s in an LHe TPC; also to have enough ionized electrons to be separated from their neighbor ions and eventually drift to the helium gas layer to generate electroluminescence, the field should be $\ge$ 10 kV/cm. A 1-meter height TPC~\footnote{A LHe TPC having the diameter and height of 1-meter can fill $\sim$ 100 kg LHe.} would therefore require 1-million volts. High voltage requires special care on HV power suppliers, cables, feed-throughs, and spark and discharge mitigations. Although challenging, these technical issues are solvable according to the review panel and the experiences of the respected LXe and LAr TPCs. We will further discuss HV in section~\ref{sec4QsRs-sub2S1S2subHV}.\\

The panel suggested two independent and complementary R\&D programs: 30 g and 10 kg LHe detectors. 
The 30 g phase represents a cylindrical apparatus with a total mass of 30 g LHe; the radius and height equal 3 cm. There will be multiple versions of this setup to test.

Cal-I: ER and NR calibrations. 

Cal-II: S2 signal optimizations.

Cal-III: SiPM testing at 4 K.

The 30 g detector program is suited to answer the initial questions concerning the fundamental responses of the liquid helium to incident particles (neutrons and gammas/electrons) and establish the optimum operating conditions. 

The 10 kg detector is also cylindrical, with both the diameter and height of $\approxeq$ 50 cm, which would be necessary to demonstrate the viability of the elevated HV levels needed for an even larger scale dark matter search experiment. It would also test other needed aspects, such as large multi-channel photodetector arrays. Furthermore, given the drift speed of the ionization signals in an LHe TPC as slow as $\sim$ 2 m/s, to constrain the overlaps in background events due to cosmic rays, the 10 kg detector should be operated underground instead of running above ground.

\subsection{Other technical challenges for an LHe TPC}\label{secProjectSubChallenges}

Beyond the challenges thankfully brought out by the panel as shown in subsection~\ref{secProjectSubProjectReview}, we received insightful comments when we presented the ALETHEIA project in variant academic events. The two challenges introduced here benefited from these communications. After analyzing carefully, we think these challenges can be mitigated significantly thanks to the especially low backgrounds of the detector. However, we have to admit that the discussion here in this subsection is very preliminary. To answer such questions robustly, we need detailed simulations based on Geant4 and robust analysis on real data. Unfortunately, neither of these is available for the time being. 

\subsubsection{Self-shielding}\label{secProjectSubChallengesSubsubSelfshielding}

Compared to xenon, helium has a smaller scattering cross-section to fast neutrons. For instance, the total cross-section for 300 keV neutrons and $^4$He is 8.0e-25 cm$^2$; it's a factor of 6 smaller than $^{131}$Xe, 4.6e-24 cm$^2$~\cite{NeutronXSBNL}. For 300 keV neutrons, the Mean Free Path (MFP) in LHe and LXe is 16.2 cm and 57.3 cm, respectively; a factor of 4 difference. As a result, an LHe TPC should have a thicker shielding layer than an LXe detector to shield the same energy neutrons out of the fiducial volume. In general, the thicker the self-shielding layer, the less the number of background events, and the smaller the fiducial mass of the detector. Given that the expected ER backgrounds of an LZ-size LHe TPC without any self-shielding are 2 orders lower than the LZ detector having $\sim$ 10 cm LXe self-shielding, as will further discuss in~\ref{secProjectSubChallengesSubsubSlowMobility}, the requirement for self-shielding on an LHe TPC is very much mitigated. We will make a trade-off to optimize the thickness of the self-shielding LHe layer.

\subsubsection{Electron's slow mobility related concerns}\label{secProjectSubChallengesSubsubSlowMobility}

As mentioned in table~\ref{tabLHeProperties}, the mobility of electrons (bubbles) in LHe at 4 K is 2$\cdot 10^{-2}$cm$^{2}$/(V$\cdot$s)~\cite{Schwarz72}, which corresponds to 2 m/s under an field of 10 kV/cm. While in LAr and LXe, the electron's velocity is around 10$^3$ m/s with $\sim$1 kV/cm~\cite{AprileDoke10}. An electron's drift velocity in an LHe TPC is roughly 2 orders slower than an LXe TPC, which raises concerns on (a) S2 events overlap and (b) insufficient exposure due to too much dead time ( i.e., drift time). However, these two concerns are not noticeable because an LHe TPC's backgrounds is 2 orders less than an LXe detector of the similar size; and TPCs are essentially triggered by these background events.

According to LZ TDR~\cite{LZTDR17}, among the total 1244 ER backgrounds, 911 are dispersed radionuclides (Rn, Kr, Ar), 255 are Neutrinos ($\nu$ - e), 67 are $^{136}Xe \rightarrow 2\nu \beta \beta$~\cite{LZTDR17}. These backgrounds are almost null in an LHe detector~\footnote{Since the number of electrons of Xe is more than 30 times more than He, so the $\nu$ - e backgrounds are ignorable in an LHe detector, a dedicated calculation will be supplied in the future.}, which means an LHe detector with the size of the LZ detector ($\sim$ 1.5-m height and diameter) could only have 11 ( =  1244 - 911 - 255 - 67) ER backgrounds, two orders less than the LZ detector. For NR backgrounds, we can project similarly based on the LZ detector. Among the total 1.2 NR background events that would register, the 0.72 NR events contributed by neutrinos to the detector can be ignored in the 1.5-m LHe detector. So, the expected NR background events would only be $\sim$ 0.5. We ignore the NR events for the analysis since 0.5 $\ll$ 11. Consequently, we expect the total number of background events in a 1.5-m size TPC is 11 with 3 years running. 

As shown in reference~\cite{BkgAnalysisLZNov2022}, the background events' rates are nearly flatly distributed among all energies up to 2700 keV. We, therefore, believe the background ratio of ALETHEIA and LZ can hold for all energy background events. That being said, the ALETHEIA trigger rate (for all energy backgrounds) would be about 11 / 1244 $\approxeq$ 1 \% of LZ, i.e., 40 Hz * 1\% = 0.4 Hz. In summary, even though the electron's mobility of an LHe TPC is 2 orders slower than an LXe detector, thanks to its 2 orders lower trigger rate, we have no reason to worry about the events overlap and dead time issues in a 1.5-m LHe TPC ($\sim$ 330 kg LHe). With 3 years of scientific running, such a detector can accumulate to 1 ton*yr exposure, reaching the cross-section of $\sim$ 10$^{-45}$cm$^2$, touching down the $^8$B neutrino floor.

Here, we just estimated the backgrounds of an LHe TPC in a very preliminary way. A concrete Geant4-based simulation is required to characterize it fully; only actual data can conclude the detector's performance.

\subsection{Not covered topics}\label{secProjectSubNotCovered}

To build an LHe TPC, many other topics should be addressed, including but not limited to: cryogenics, electronics and DAQ, recycling and purification, screening, and simulation. However, considering the project is still at an early stage, we do not think it is appropriate to introduce in this paper. Instead, we will supply them in the further document(s). 


\section{The design of the ALETHEIA detectors}\label{secDesignALETHEIA}

The ALETHEIA project has been inspired by the communities of liquid helium and dark matter direct detection (especially the experiments implementing TPCs). Liquid helium has been employed (or has tried to employ) in variant research, for instance, EDM~\cite{Huffman2000, Ito2016}, Solar neutrinos~\cite{HOERN1}, and even dark matter~\cite{HOERN88, HOERN13, GuoMckinsey13}. Though most of these liquid helium detectors utilize superfluid helium, which is different from the ALETHEIA detector, their researches are very helpful for our calculations and simulations. In addition, the respected LXe and LAr TPCs experiments present direct experiences for us to design our TPC. Finally, theoretical physicists also thankfully contributed to the project in several ways, such as project reviews and insightful private comments/discussions. As a result, we do not consider that the ALETHEIA project has been contributed by a single person or collaboration alone. Instead, it represents the wisdom of the related communities, provided it can grow to be capable of helping answer one of the most critical questions in fundamental physics today: the nature of dark matter.

\subsection{The conceptual design of the ALETHEIA detectors}\label{sec3subsec1}

\begin{figure}[!t]
	\centering
        \includegraphics[width=5.0in, angle = 0]{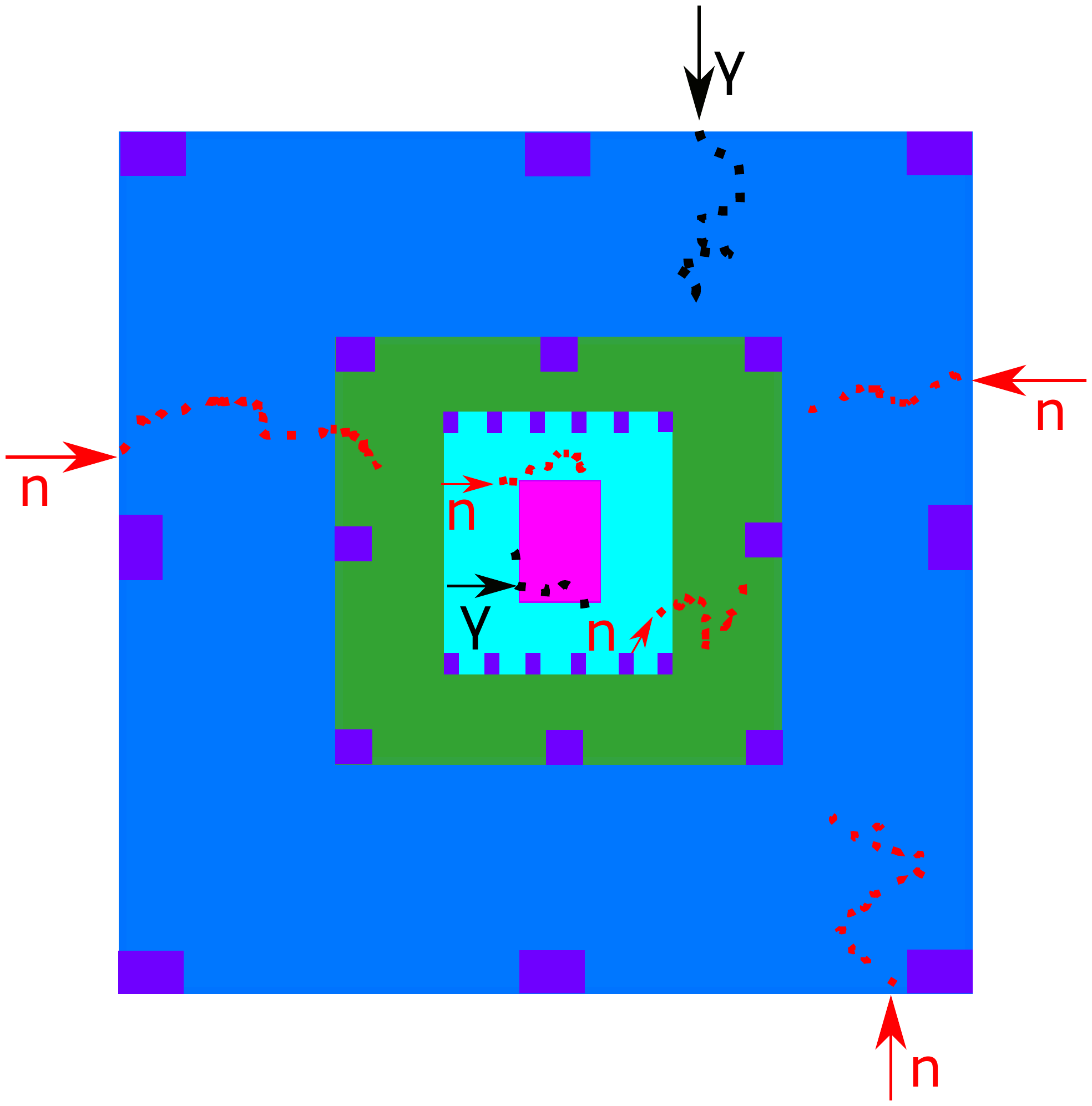}   
	\caption{The schematic drawing of the ALETHEIA detector (not to scale). From outside to inside: The light blue area represents the water tank surrounding the whole detector system, with a diameter of a few meters; the eight purple rectangles on the edge of the water tank are the PMTs to detect background signals inside of the water tank; the dark green is the Gd-doped liquid scintillator veto, with the thickness of $\sim$ half-meter; the eight purple rectangles at the edge of veto (green area) are PMTs to detect scintillation insides of the veto detector; the cyan area is the active volume of the TPC, filled with liquid helium; the total twelve purple rectangles on the top and bottom of the active volume (cyan area) represent the SiPMs to detect scintillation and electroluminescence; the magenta region represents the fiducial volume of the TPC where extremely low backgrounds are expected there. The red and black dots represent the tracks induced by background neutrons and $\gamma$s, respectively.}
	\label{FigSchematicALETHEIA}
\end{figure}

Figure~\ref{FigSchematicALETHEIA} shows the schematic drawing of such a detector. As shown in the figure, the core of the ALETHEIA experiment is a dual-phase liquid helium TPC (cyan). The TPC central magenta area represents the fiducial volume where extremely low or zero background is expected. On the top and bottom of the TPC are twelve purple rectangles, which represent SiPMs. The TPC was surrounded by a Gd-doped scintillator detector (green), which acts as a veto. The utmost is a water tank (blue) a few meters in diameter to shield neutrons and gammas outside the detector system. All of the parts in the figure are not to scale.

For neutrons that come from outside the water tank, a few meters of water should be thick enough to thermalize most, if not all, of them. The $\sim$ half-meter Gd-doped liquid scintillator would capture thermalized neutrons. Neutrons originating from the TPC inside could be identified by the feature of multiple hits in the TPC and (or) liquid scintillator. The reason is that the interaction between neutron and helium nuclei is strong interaction. As a comparison, the hypothetical WIMPs signals would have only one hit because the cross-section between WIMPs and helium nuclei is weak interaction or more minor. For $\gamma$s from outside of the water, the water tank can block them from entering the central detector; for $\gamma$s from inside of the TPC, S1/S2, or PSD, or a hybrid analysis combined these two analyses could, in principle, discriminate them from nuclear recoils induced by a neutron or WIMPs.

\section{LHe TPCs' R\&D plan}\label{sec4QsRs}

The R\&D plan is driven by the three analysis methods, PSD, S1/S2, and S2O, which will possibly all be implemented in ALETHEIA. We first list some of the 4 K LHe properties in table~\ref{tabLHeProperties}. We will cite them frequently in the following sections. 

\begin{table}[h]
\begin{minipage}{328pt}
\caption{Some of related 4 K  LHe properties}\label{tabLHeProperties}%
\begin{tabular}{@{}llll@{}}
\hline
\toprule
Property & Value  & Reference \\
\midrule
      Ionization energy	 (E$_i$)			& 24.6 eV 						&\cite{Jesse55}  \\
      \vtop{\hbox{\strut Average energy to produce }\hbox{\strut an electron-ion pair (W)}} 	& \vtop{\hbox{\strut ~}\hbox{\strut 42.3 eV}}  		&\vtop{\hbox{\strut ~}\hbox{\strut \cite{Jesse55} }}   \\
      Excitation energy (E$_e$)				&20.6 eV			&\cite{Smirnov82} \\
      \vtop{\hbox{\strut Lifetime of prompt singlet scintillation,}\hbox{\strut produced by electron-ion pairs recombination}} 	& \vtop{\hbox{\strut ~}\hbox{\strut $<$ 10 ns}}  		&\vtop{\hbox{\strut ~}\hbox{\strut \cite{Keto72} }}   \\
      \vtop{\hbox{\strut Lifetime of ``delayed'' singlet scintillation,}\hbox{\strut produced by excitation: He$^*$+He $\rightarrow$ He$_2^*$ (singlet) }} 	& \vtop{\hbox{\strut ~}\hbox{\strut $\sim$ 1.6 $\mu$s}}  		&\vtop{\hbox{\strut ~}\hbox{\strut \cite{McKinsey03} }}   \\
      \vtop{\hbox{\strut Lifetime of triplet scintillation,}\hbox{\strut produced by both excitation and ionization}} 	& \vtop{\hbox{\strut ~}\hbox{\strut  13$\pm$2 s}}  		&\vtop{\hbox{\strut ~}\hbox{\strut \cite{McKinsey99} }}   \\
	\vtop{\hbox{\strut Calculated G-values on 100 keV electrons for:}\hbox{\strut ionization, singlet, triplet }}    	&\vtop{\hbox{\strut ~}\hbox{\strut 2.27, 0.85, 0.17}}  	&\vtop{\hbox{\strut }\hbox{\strut \cite{ERLHeCalculation74}} }\\
	\vtop{\hbox{\strut Calculated G-values on 20 keV neutrons for:}\hbox{\strut ionization, singlet, triplet }}    	&\vtop{\hbox{\strut ~}\hbox{\strut 1.30, 1.83, 0.0004}}  	&\vtop{\hbox{\strut }\hbox{\strut \cite{ERLHeCalculation76}} }\\
	Electron's  mobility@ 4 K		& 2$\cdot 10^{-2}$ cm$^{2}$/(V$\cdot$s)  	&\cite{Schwarz72}\\
	Electron's  velocity@ 4 K, 10 kV/cm		& 2 m/s  	&\cite{Schwarz72}\\
      \vtop{\hbox{\strut Breakdown voltage for helium gas at a}\hbox{\strut  saturated vapor pressure (4.2 K, 1 atm)}} 	& \vtop{\hbox{\strut ~}\hbox{\strut  $\sim$100 kV/cm }}  		&\vtop{\hbox{\strut ~}\hbox{\strut \cite{GuoMckinsey13} }}   \\

      \vtop{\hbox{\strut Breakdown voltage for liquid helium}\hbox{\strut  (1.2 - 4.2 K)}} 	& \vtop{\hbox{\strut ~}\hbox{\strut  $\sim$1 MV/cm }}  		&\vtop{\hbox{\strut ~}\hbox{\strut \cite{Blank60} }}   \\
\botrule
\end{tabular}
\end{minipage}
\end{table}

\subsection{The ``input'' of an LHe detector: particles hint LHe, }\label{sec4QsRsSubsec1Input}

Energetic particles passing through a medium of LHe will deposit part or all of its kinetic energy. If the incident particle is an electron (or $\gamma$), it interacts with the electrons of helium atoms via electromagnetic interaction. The atoms then either get ionized or excited or both. If the incident particle is a neutron, it interacts with helium atoms with strong interaction. As a result, the helium atoms get recoiled and become moving particles, $\alpha$s. The $\alpha$ particles further interact with the electrons of helium atoms electromagnetically, i.e., ionize and excite helium atoms along their trajectories. 

Although electrons and $\alpha$ particles interact with helium atoms via electromagnetic interaction, the charge densities of ions and electrons are different. As mentioned in reference~\cite{McKinsey03}, for $\sim$ MeV electrons, energy deposition is 50 eV$\mu$m$^{-1}$; while for $\alpha$ particles, energy deposition is 2.5 $\times$10$^4$ eV$\mu$m$^{-1}$. For liquid xenon, similar results were obtained with simulation~\cite{DahlPhDThesis09}.  
The charge density depends on the particle's stopping power, or ``dE/dx''~\cite{McKinsey03, AprileDokeReview2009}. According to the \textit{Bethe formula}, for low energy incident particles ( $v \ll c$, where $v$ is the velocity of particles, $c$ is the speed of light.), $dE/dx \sim \propto 1/v$, while for the same kinetic energy $\alpha$s and electrons (100 keV for instance), the velocity of electrons is roughly 2 orders faster than $\alpha$s, as a result, $dE/dx$ of electrons is roughly 2 orders smaller than $\alpha$s. Consequently, the charge densities for ER/NR have 2 orders difference. 

Moreover, the geometry and separation of the tracks induced by ER and NR are different in LHe: ER events are ``small dots'' in shape and separated on average 500 nm; NR events are cylindrical and separated, on average, 1 nm. Considering the distance between the ion and its separated electrons is roughly 20 nm, ER events are well separated (500 nm $ \gg $ 20 nm), and the recombination is geminate~\cite{Onsager38}, meaning the recombined electron-ion pair is the one being separated moments ago. In comparison, NR events are heavily overlapped between individual ionization events (1 nm $ \ll $ 20 nm)~\cite{McKinsey03}, the recombination is columnar~\cite{Jaffe1913, Kramers52}, meaning the recombined electron-ion pair is not necessarily the one just separated. 

\subsection{The ``output'' of an LHe detector: Electrons and scintillations}\label{sec4QsRsSubsec2Output}

An energetic particle ($\alpha$, $\beta$, $\gamma$, or neutron) will leave electron-ion pairs and excited helium atoms in LHe. The recombination of ionized electrons and ions and the de-excitation of excited atoms will generate scintillations. A strong enough external electric field can separate some or all ionized pairs. The separated pair can not recombine to generate scintillation anymore, but both the electrons and irons can drift along the electric field; drifting electrons will be further accelerated at a $\sim$ 1-cm thick helium gas layer under a local HV field to generate electroluminescence (i.e., S2 signals). For specific deposited energy, the yield of ionized electron-ion pairs and scintillations can be calculated~\cite{ERLHeCalculation74, ERLHeCalculation76}. Increasing the external field will get greater S2 but result in smaller S1 simultaneously. However, the electric field will not affect the scintillation generated by excited atoms and their following de-excitation.

Three scintillation components exist in LHe: 10 ns, 1.6 $\mu s$, and 13 s. The 10 ns prompt scintillation have a lifetime of $<$ 10 ns, which are resulted from the decay of excited singlets through recombinations, as shown in Eq.(\ref{eq1}).

\begin{equation}
\text{He}^*_{2}( \text{A} ^{1} \sum_{\mu}^{+}) \underrightarrow { ~~ 10 ~ns ~~} 2 \text{He}(1^{1} \text{S} ) + h\nu \label{eq1}
\end{equation}

The 1.6 $\mu$s originated from the decay of an excited atom ($\text{He}^*(_2^1\text{S})$) and ground helium atom ($\text{He}(_2^1\text{S})$) formed dimer, as shown in Eq.(\ref{eq2}).
 
 \begin{align}
 \text{He}^*(_2^1\text{S}) +\text{He}(_2^1\text{S}) \underrightarrow { ~~ 1.6 ~\mu s ~~}  &\text{He}_2^* ( \text{A} ^{1} \sum_{\mu}^{+}) \nonumber \\
&\text{He}_2^* (\text{A} ^{1} \sum_{\mu}^{+}) \underrightarrow { ~~ 10 ~ns ~~} 2 \text{He}(1^{1} \text{S} ) + h\nu  \label{eq2}
\end{align}

The 10 ns and the 1.6 $\mu s$ fluorescence are due to singlets decay. However, the  13 s scintillation is phosphorescence, resulting from triplets decay. The lifetime of phosphorescence is much longer because the decay flips the excited molecular dimer's spin from 1 to 0~\cite{McKinsey99}. 

Be aware, please, that the mechanism of 1.6 $\mu $s scintillation production in LHe is different from in LAr. In LHe, the 1.6 $\mu $s is due to two processes: the relatively longer process (1.6 $\mu $s) of the form of an excited \textit{singlet} dimer and the shorter decay (10 ns), as shown in Eq.(\ref{eq2}). In LAr, the 1.6 $\mu $s component is the decay from an excited \textit{triplet} to a ground singlet~\cite{Hitachi83}. 

In the theoretical LHe community, many efforts have been made to understand the output of incident particles. Reference~\cite{ERLHeCalculation74} calculated the G-values for 100 keV electrons hitting on helium gas. Reference~\cite{ERLHeCalculation76} compared the G-values for $\alpha$, $\beta$, and $\gamma$; it also calculated the G-values for $\alpha$s between 20 keV and 8 MeV. The most relevant calculating results are: for 100 keV electrons, the G-values for ionization, singlet, and triplet are 2.27, 0.85, and 0.17, respectively; while for 20 keV $\alpha$s, the values are 1.3, 1.83, and 0.004, as showed in table~\ref{tabLHeProperties}. That being said, assuming the incident particle is a 100 keV electron, for every produced electron-ion pair, there will be 0.45 scintillation photons composed of 0.37 singlet and 0.08 triplet. In the case of 20 keV $\alpha$s, each ionized pair corresponds to 1.4 photons decayed from singlets and ignorable triplet scintillation. 

\begin{figure}[!t]	 
	\centering
        \includegraphics[width=4.5in, angle = 0]{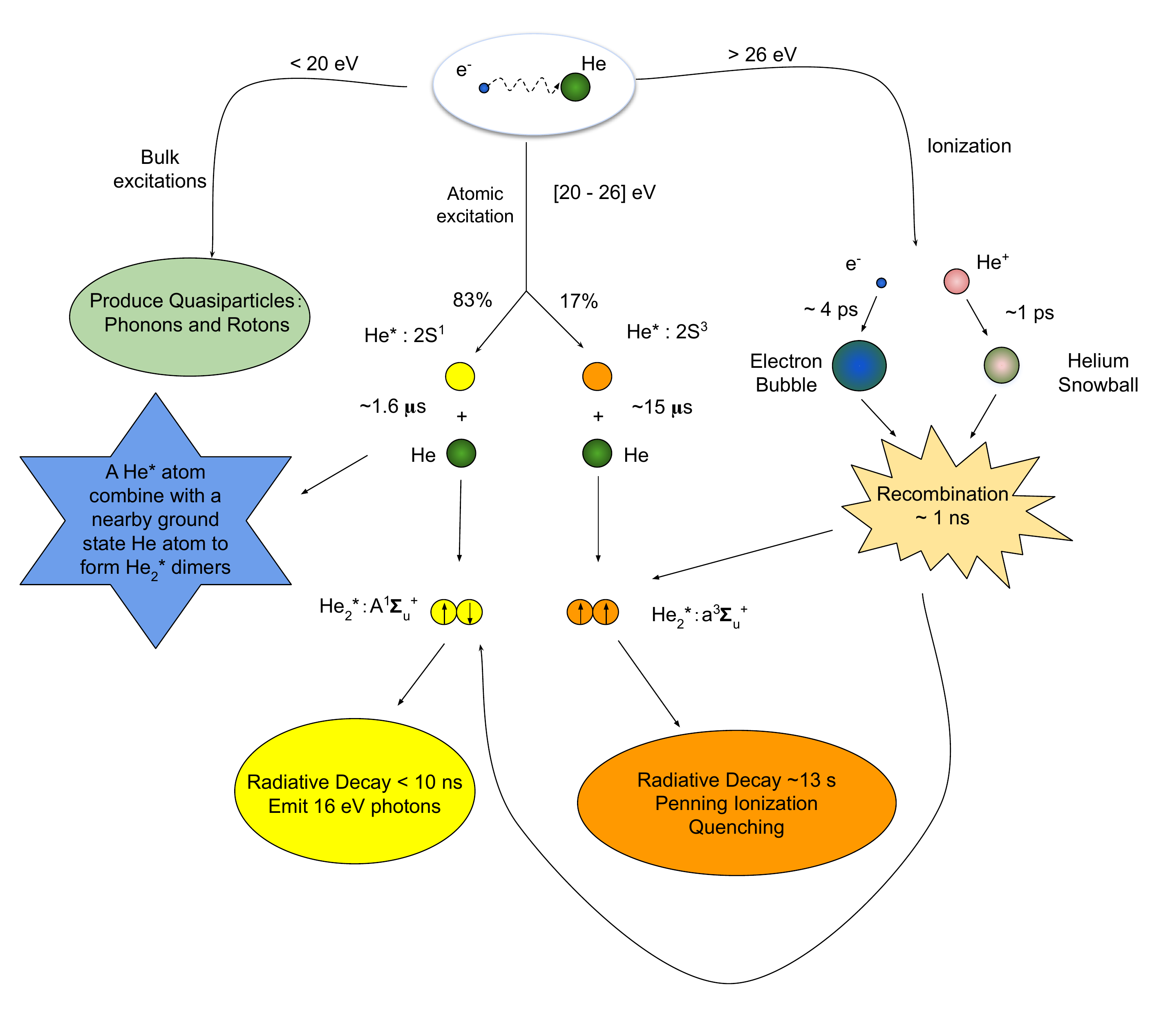}
	\caption{The schematic drawing describes the mechanism of scintillation production in LHe when hitting by 100 keV electons~\cite{ERLHeCalculation74}.}\label{fig_LHeScintillationProdMechnism}
\end{figure}

Fig.~\ref{fig_LHeScintillationProdMechnism} shows the scintillation production mechanism in LHe schematically with 100 keV incident electrons. The figure is inspired by reference~\cite{DaifeiJinPhDThesisBrown12}.

\subsection{LHe scintillation yield and its application to PSD}\label{sec4QsRs-sub1ScintYield}

Essentially, the PSD technique in LAr experiments employs the different time features of scintillations to discriminate ER/NR events; specifically, the different intensity ratio of scintillation produced by singlet and triplet excimers, ratio = (10 ns scintillation intensity) / (1.6 $\mu s$ scintillation intensity). For LAr, the lifetime of singlet and triplet is $\sim$ 10 ns and $\sim ~1~ \mu$s, respectively. The two orders difference in a lifetime is proven to be able to apply PSD for ER/NR in DarkSide-50 and DEAP. For LXe, the difference of the lifetime for singlet and triplet is only one order~\cite{Hitachi83}. Therefore, LXe experiments have not implemented the PSD as a primary technology to discriminate ER from NR events; instead, they use the S1/S2 analysis. If the PSD analysis can not perform well enough in an LHe TPC for whatever reason(s), ALETHEIA will mainly rely on the S1/S2 analysis.

\subsubsection{Accomplished researches on LHe scintillation}\label{sec4QsRs-sub1ScintYieldSubAccReser}

McKinsey and colleagues measured the lifetime of triplets in LHe to be $\sim$13 s~\cite{McKinsey99}; they also confirmed the existence of 1.6 $\mu$s scintillation in LHe~\cite{McKinsey03} with 5.3 MeV $\alpha$ ($^{210}$Po), 364 keV $\beta$ ($^{113}$Sn), and proton and $^3$H (produced by $^{3}$He capturing neutron). Furthermore, they found the 1.6 $\mu$s component is relatively weaker than the prompt singlet ($< 10 $ ns) and triplet scintillation (13 s) when irradiated by $\beta$ compared to $\alpha$ and neutrons. This is because $\alpha$s induced ionizations will take Penning interaction, producing extra electrons and ions. The recombination of these particles can generate 1.6 $\mu$s component.
McKinsey and his group members also tested scintillation on a 1.0 cm scale LHe apparatus with 0.546 MeV $^{90}$Sr and 2.28 MeV $^{90}$Y. They explicitly observed that ``prompt scintillation pulse'' and ``Afterpulse scintillations'' represent 10 ns and 1.6 $\mu$s scintillation, respectively~\cite{GuoJinst12}.  

Using 5.5 MeV $\alpha$ particles ($^{241}$Am), Ito et al. measured the scintillation yield change along with an external field; they found (a) prompt scintillation yield reduced by 15\% at 45 kV/cm and (b) the electric field has a stronger effect on the yield of 1.6 $\mu$s component than the prompt one~\cite{Ito2012}~\footnote{The result of ``the electric field has a stronger effect on the yield of the 1.6 $\mu$s component than the prompt one'' is confusing at first glance. Because the 1.6 $\mu$s component is supposed to result from excitation and de-excitation, do not relate to ionization. The fact is that there is an additional source of the 1.6 $\mu$s component when $\alpha$s hit LHe, which is field sensitive. The paper~\cite{Ito2012} interpreted that compared to 100 keV electrons, 5.5 MeV $\alpha$s generate a much higher density of ionized electrons and irons in LHe; before recombinations happen, triplets will take Penning interaction to produce a significant number of electrons and irons. These Penning interactions produced electrons-irons would go through an extra time of separation by the field, reducing combinations and, therefore, the production of 1.6 $\mu$s component scintillation.}. With another apparatus, they measured the prompt scintillation produced by 364 keV $\beta$ ($^{113}$Sn) and found a 42\% reduction when a 40 kV/cm field was applied to the detector~\cite{Phan20}. However, for the interesting NR energy interval of the ALETHEIA, [ 0.4 keV$_{nr}$, 10 keV$_{nr}$], no similar measurement has been performed yet. The ALETHEIA project would launch such calibrations with $\sim$ 10s keV mono-energetic accelerator neutrons from the $^{45}$Sc(p,n) interaction~\cite{LAMIRAND20101116} and 17 keV mean energy electrons from $^{63}$Ni~\cite{Ni63SourceIAEA}, respectively.  

\subsubsection{Applying the S2O analysis for DM searches}\label{sec4QsRs-sub2S2OSubS2ODMSearches}

The S2O analysis of noble gas TPCs could be complementary to the PSD and S1/S2. Compared to the S1/S2, the S2O analysis loses the position sensitivity of a TPC; therefore, it is not competitive in terms of background events constraint. However, the S2O analysis is still convincing as long as the observed events are consistent with anticipated backgrounds for selected datasets; moreover, it can extend the TPC's sensitivity roughly one order lower on WIMPs mass than the S1/S2 and S2O according to DarkSide-50~\cite{DarkSide502018S2O} and XENON-1T~\cite{XENON1TS2O19}.

The strategies of the S2O analysis are not the same for the two experiments. For DarkSide-50, as mentioned in reference~\cite{DarkSide502018S2O}, to reach the lowest possible S2 signals, the fiducial volume can not be reconstructed with the usual algorithm due to low photoelectron statistics for S2, the fiducial region for the S2O analysis is in the $x-y$ plane by only accepting events where the largest S2 signal is recorded in one of the seven central top-array PMTs. For XENON-1T~\cite{XENON1TS2O19}, they used 30\% of Science Run (SR1) data as training data to determine event selections. Limits settings are computed using only the remaining 70\% data, which was not examined until the analysis was fixed. 

Our preliminary plan is to implement the 10 ns and the 1.6 $\mu s$ scintillation for PSD analysis, though the 13 s channel is also possible. We will investigate this with dedicated calibrations in an underground lab.
Without data on hands, it is difficult to project the sensitivity of ALETHEIA for the S2O analysis for the moment. 
Prior, a few pre-R\&D should be launched: wavelength shifter and photon sensors. 

LHe scintillation's wavelength peaks at 80 nm (or 16 eV). The wavelength is out of the detectable range of current commercially available photon sensors. So a wavelength shifter like tetraphenyl butadiene (TPB) is required to shift the photons to $\sim$450 nm~\cite{Lippincott12}. Reference~\cite{Benson18} measured the efficiency of TPB coating on an acrylic substrate at room temperature in a vacuum vessel. They obtained $\sim$ 30\% wavelength shifter efficiency for 80 nm light. The reference also predicted a higher efficiency than the measured one when the TPB layer works at LHe temperature. We have successfully coated a $\sim 3 ~\mu $m TPB layer on the whole inner walls of a 10-cm scale cylindrical PTFE vessel~\cite{TPBCoutingALETHEIA22}, as will be further discussed in section~\ref{secProgressALETHEIA}.

In general, the smaller the WIMPs' mass, the lower the kinetic energy of WIMPs, the lower the energy for excitation and ionization, and the less the number of scintillation photons. Therefore, considering ALETHEIA's ROI is low-mass WIMPs, we will choose SiPMs as our photon detectors, thanks to the higher detection efficiency than other photon sensors, such as PMTs (Photon Multipliers). 

Surprisingly, certain types of Hamamatsu SiPMs, which were not initially designed to work at LHe temperature, turned out to be functional at such low temperatures~\cite{Cardini14, Iwai19}, or even 200 mK~\cite{ZhangSiPM2022}. Moreover, the relative PDE (Photon Detection Efficiency) near LHe temperature (5 K and 6.5 K) is roughly to be 70\% of room temperature according to the two independent tests~\cite{Cardini14, Iwai19}. We have tested more than 10 pieces of FBK SiPMs at $\sim$ 4 K temperature and successfully demonstrated that FBK SiPMs are capable of working at LHe temperature, as will be discussed in section~\ref{secProgressALETHEIASubTestingSiPMs}.

\subsection{S1/S2 analysis in LHe}\label{sec4QsRs-sub2S1S2}

As already mentioned, the S1/S2 analysis plays a key role for ALETHEIA. Consequently, electron detection would become the most critical part of the R\&D program. We have several approaches to achieve electron detection as the following.

(a) Extracting electrons into GHe (Gaseous Helium) from LHe, producing proportional scintillation light as other two-phase xenon and argon TPCs. 

(b) GEM-based electron detection in the LHe or GHe. Reference~\cite{Erdal15} demonstrated this technology in an LXe apparatus, with a thick GEM to multiply electrons in a GXe bubble produced by a heated wire beneath. 

(c), Electroluminescence due to electrons entering very high fields surrounding thin wires immersed in the LHe. This method's mechanism is similar to a gas-proportional tube where electrons' avalanche happens as long as electrons approach the anode of the tube.

(d), Charge amplification using small-scale structures, such as Micromegas.

\subsubsection{S1/S2 for ALETHEIA}\label{sec4QsRs-sub2S1S2subALETHEIA}

LXe and LAr dark matter experiments have demonstrated that the S1/S2 technique can discriminate ER/NR events and define a fiducial volume. In principle, the same analysis methods could be transplanted into ALETHEIA because the difference in charge density of ER and NR events in LXe and LAr also holds in LHe. The ALETHEIA TPC's design would follow many experiences of the DarkSide-50 and DarkSide-20k detectors. Typical light transportation in an LHe TPC is like this: the 80 nm scintillation will first be shifted to $\sim$ 450 nm with $\sim 3~ \mu$m TPB, then go through a $\sim$15 nm ITO coated acrylic plate (1.5 cm), finally detected by SiPMs~\footnote{The facing LHe side of the 1.5 cm-thick acrylic plate first coated with a $\sim$ 15 nm ITO layer, the $\sim 3~ \mu$m TPB layer is coated on the ITO.}. 

Although in the ALETHEIA's interesting recoil energy range, [0.4 keV$_{nr}$, 10 keV$_{nr}$], there exist no experimental data verified models for calculation, we can still do calculations with reasonable estimations. As an example, we compare the number of Photoelectrons (Phe) of S1 and S2 for 1.0 keV$_{nr}$ NR and its equivalent quenched energy 0.2 keV$_{ee}$ ER~\cite{Santos08}.

For NR, according to Seidel's calculations~\cite{SeidelPKUTalk2019}, the 1.0 keV$_{nr}$ would generate $\sim$ 40 prompt scintillation photons and roughly 2.0 primary electrons.  Considering the 30\% efficiency of a $\sim 3~ \mu$m TPB to convert 80 nm scintillation into 450 nm light~\cite{Benson18}, and the 40\% photon detection efficiency of FBK SiPM~\cite{DarkSide20k17}, the 90\% total efficiency for the 15 nm ITO layer and 1.5 cm acrylic to transmit visible light~\cite{AgnesDarkSide502015, DarkSide502018}, the detected number of Phe would be 40 * 30\% * 40\% * 90\% $\approxeq$ 4.3. The 4.3 Phe is the S1 signal of an LHe TPC corresponding to 1.0 keV$_{nr}$. The 2.0 electron is the source of S2. If the field is 10 kV/cm, the probability of being separated is 50\%~\cite{Phan20}, as will be further discussed below. Assuming the efficiency of the electron extracting from helium liquid to gas is 80\%, the total detection efficiency for the electron is 40\% ( = 50\% * 80\%). Assuming the amplification factor of S2 is 800 Phe/e (the same as LZ, \cite{LZTDR17}), the S2 size is 640 Phe ( = 2.0 e * 40\% * 800 Phe/e).

For ER, the 0.2 keV$_{ee}$ would generate 3.0 prompt scintillation photons and 4 primary electrons~\cite{SeidelPKUTalk2019}. Following the same algorithms, the S1 and S2 signals are 0.3 Phe ( = 3 * 30\% * 40\% * 90\%) and 1280 Phe ( = 4 e * 40\% * 800 Phe/e), respectively.

As summarized in table~\ref{tabCompERNRp4keV}, for 1.0 keV$_{nr}$ NR events, the ALETHEIA TPC could detect a 4.3-Phe size S1 and a 640-Phe size S2 on average; for 0.2 keV$_{ee}$ ER events, the S1 and S2 are 0.3 Phe and 1280 Phe, respectively. Under the plan of Log$_{10}$(S2) V.S. S1, the S1 signals have a factor of 14.3  ( = 4.3 / 0.3) difference for ER and NR events, S2 signals exist a factor of 0.9 (Log$_{10}$(640) / Log$_{10}$(1280)) difference, which should be enough to discriminate each other.

In the above calculation on S1, we did not consider the effects of temperature and external field. Combining these two factors, the yield likely will decrease $\sim$ 10 - 20\% for ER and NR events in a 4 K LHe TPC at 10 kV/cm field, according to the extrapolation from references~\cite{Ito2012, Phan20}.

\begin{table}[h]
\begin{minipage}{328pt}
\caption{Signal Comparison on 0.2 keV$_{ee}$ ER and 1.0 keV$_{nr}$ NR events in a LHe TPC}\label{tabCompERNRp4keV}
\begin{tabular}{@{}cccccc@{}}
\toprule
Recoil types & Energy  & S1 (Phe) & S2 (Phe) &S1 normalize &Log$_{10}$(S2) normalize\\
\midrule
      ER	&0.2 keV$_{ee}$ 	&0.3		&1280 	&1.0		&1.0\\
      NR	&1.0 keV$_{nr}$ 	&4.3 		&640  	&14.3		&0.9\\
\botrule
\end{tabular}
\end{minipage}
\end{table}

Some experimental efforts have been made to calibrate the S1 and S2. For example, reference~\cite {SethumadhavanPhDThesis07} tested the S2 currents with a 1-cm scale LHe cell. They found that a significant electroluminescence current can generate at the 1 atm helium gas under the voltage of $\sim$ 500 V/cm. This test demonstrated that a dual-phase TPC filled with helium could generate electroluminescence (or S2) signals, making the ALETHEIA's S1/S2 and S2O analysis viable.
Reference~\cite{Seidel14} indicates that at an external electric field of [10 - 50] kV/cm, $\alpha$s and $\beta$s events do not have the same fraction of elections being collected, $\beta$s is a factor of three greater than $\alpha$s, which could be implemented in the S1/S2 analysis for ER/NR events discrimination. Ito et al. studied the relationship between scintillation yield and external field with 5.5 MeV $\alpha$s and 364 keV $\beta$s. They observed prompt scintillation reduction of 15\% at 45 kV/cm and 42\% at 40 kV/cm for $\alpha$s and $\beta$s~\cite{Ito2012, Phan20}, respectively; which is factor of $\sim$3 difference in reduction for $\alpha$s and $\beta$s. The reduced scintillation should show up as the increased electrons. The consistence of measured results~\cite{Ito2012, Phan20} and calculated ones~\cite{Seidel14} justified the hypothesis. Simulations in reference~\cite{GuoMckinsey13} showed a good ER / NR discrimination in LHe for ionization energy down to 10 keV$_{\text{ee}}$ with the drift field of 10 kV/cm. 

The higher the field, the greater the S2 signal, and the smaller the S1 signal. For an LHe TPC, the preliminary drift voltage would be 10 kV/cm, much higher than LUX's 170 V/cm, LZ's 310 V/cm, and DarkSide-20's 200 V/cm. Even at such a high field, the electron collecting ratio in LHe is calculated to be 40\%~\cite{Seidel14}, though the latest measurements~\cite{Phan20} showed a greater collection efficiency, $\sim$ 50\% at 10 kV/cm.  

Nevertheless, one of the critical R\&D programs for the ALETHEIA project is developing a safe and stable HV system up to 500 kV or higher, as will be discussed in the following subsection~\ref{sec4QsRs-sub2S1S2subHV}.

\subsubsection{High voltage in the ALETHEIA}\label{sec4QsRs-sub2S1S2subHV}

As mentioned above, an electric field equal to or larger than 10 kV/cm is needed to achieve sufficient charge collection efficiency. Ito and his colleagues at LANL (Los Alamos National Laboratory) have successfully applied $>$ 100 kV/cm in a volume between 12 cm diameter electrodes separated by 1 cm~\cite{Ito16}, which indicates applying a 100 kV electrical potential into liquid helium from an external power supply component is viable. As indicated in table \ref{tabLHeProperties}, the breakdown voltage for LHe and 1 atm helium gas are 1 MV/cm and 100 kV/cm, respectively. When an LHe TPC works, local helium gas bubbles may exist inside the bulk due to heating from SiPMs or other parts. Therefore, a higher than 100 kV/cm voltage might cause local breakdowns. To ensure no breakdown, we should not apply a voltage exceeding 100 kV/cm to the detector.

To deliver 500 kV or higher voltage into the LHe TPC, we need a DC power supply, cables, feedthroughs, and measures for spark and discharge mitigation inside the LHe detector. Reference~\cite{Cantini17} reported that the Heinzinger company could provide a 300 kV DC power supply and cables. However, to have a higher HV, generating HV directly inside the liquid, using Cavallo's multiplier~\cite{Clayton18}, is likely necessary. 

As to the HV feedthrough, our preliminary plan is not to have it immersed in a cryogenic environment, as described in reference~\cite{Cantini17}; instead, we will choose the solution of the LZ TPC~\cite{LZTDR17} where the feedthrough is at Room Temperature (RT); the anode, isolation layer, and braid of the cable are all the same material (polyethylene) but have different doping components. A ``umbilical'' goes through the water tank and delivers HV down to the TPC. There are a few reasons for such an option: (a) it is much easier to purchase a feedthrough working at RT than doing at 4 K; (b) it is also more reliable since if the working temperature is at RT, no worries on the shrinking coefficients for the different materials of the feedthrough along with cooling down and warming up; (c) very limited radioactivities contributed from the feedthrough since it is outside of the water tank. 

Often, the highest field appears at the end of an HV cable where the anode and braid have the shortest distance. To mitigate possible spark and discharge, we should implement a field-grading structure at the end of the HV cable~\cite{LZTDR17}. The structure's purpose is to increase the distance between the anode and the ground braid by 10 times or so. Take 500 kV as an example, assuming the distance between the anode and the braid of an HV cable is 2 cm; without the structure, the field would be 250 kV/cm at the end of the cable. This field is 2.5 times the 1 atm helium gas breakdown voltage (100 kV/cm). With such a structure, however, the distance can be enlarged to 20 cm or more. As a result, the field could decrease to 25 kV/cm or below, which is safe for 1 atm helium gas.

\section{The progress of ALETHEIA so far}\label{secProgressALETHEIA}

The ALETHEIA project has officially launched in the summer of 2020. We have made significant progress since then. 

\subsection{The R\&D of the 30g-V1 LHe detector}\label{secProgressALETHEIASub30gV1}

\begin{figure}[!t]	 
	\centering
    \includegraphics[scale=0.2, angle = 0]{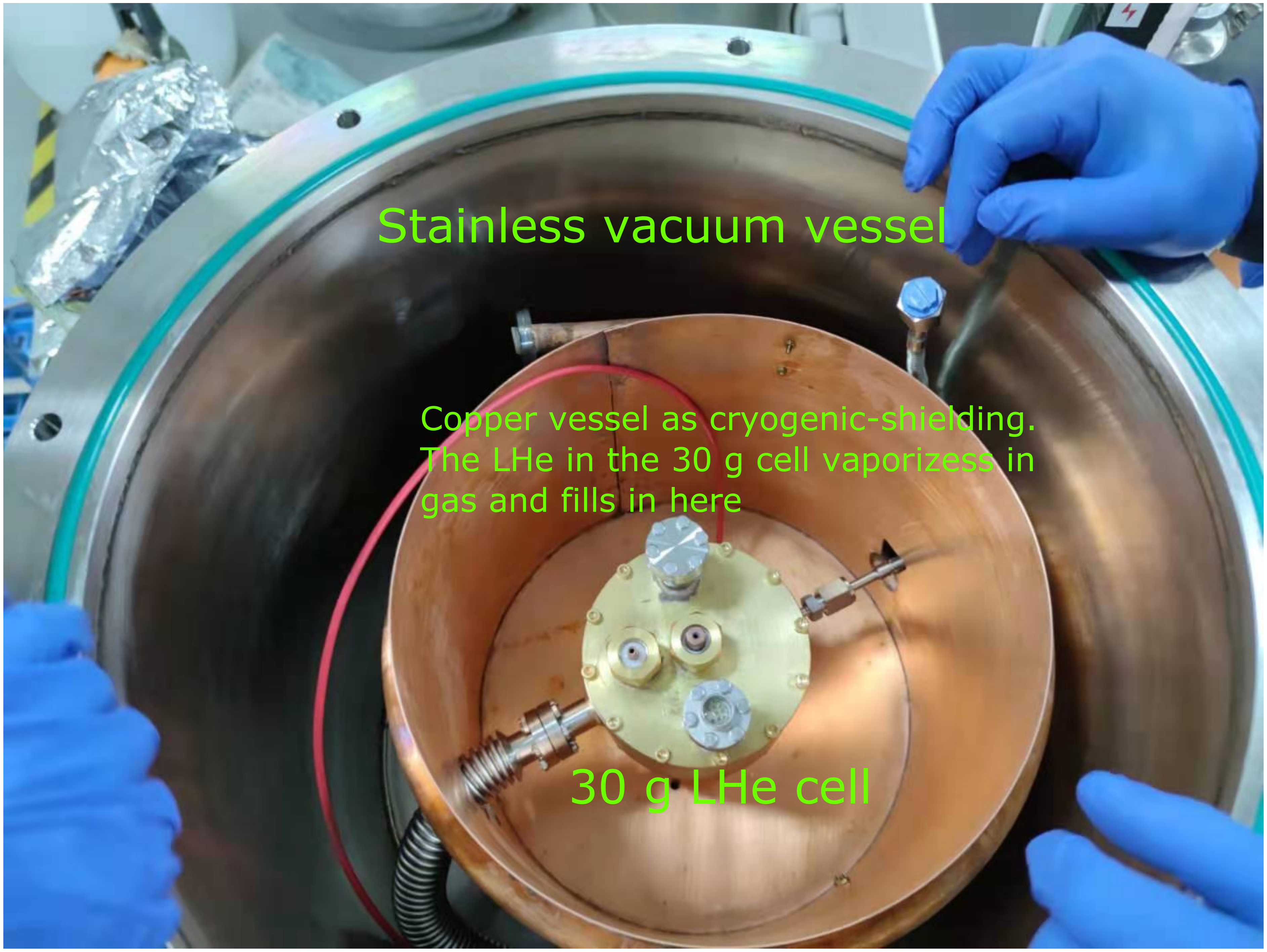}
	\caption{The inside of the 30g-V1 LHe detector (top view).}\label{fig_30gLHeCellAssembly}
\end{figure}

We built our first 30 g LHe cell at CIAE in Beijing, China, in 2021, as shown in Fig.~\ref{fig_30gLHeCellAssembly}. The primary purpose of this detector is to gain experience in building an apparatus capable of cooling down to LHe temperature ($\sim$ 4.5 K), make sure an external DC HV could be applied to the detector, and the dark current of the detector under HV should be small enough. The R\&D of the detector is proved to be very helpful to the design and assembly of the 30g-V2 detector.

\begin{figure}[!t]	 
	\centering
    \includegraphics[width=4.0in, angle = 0]{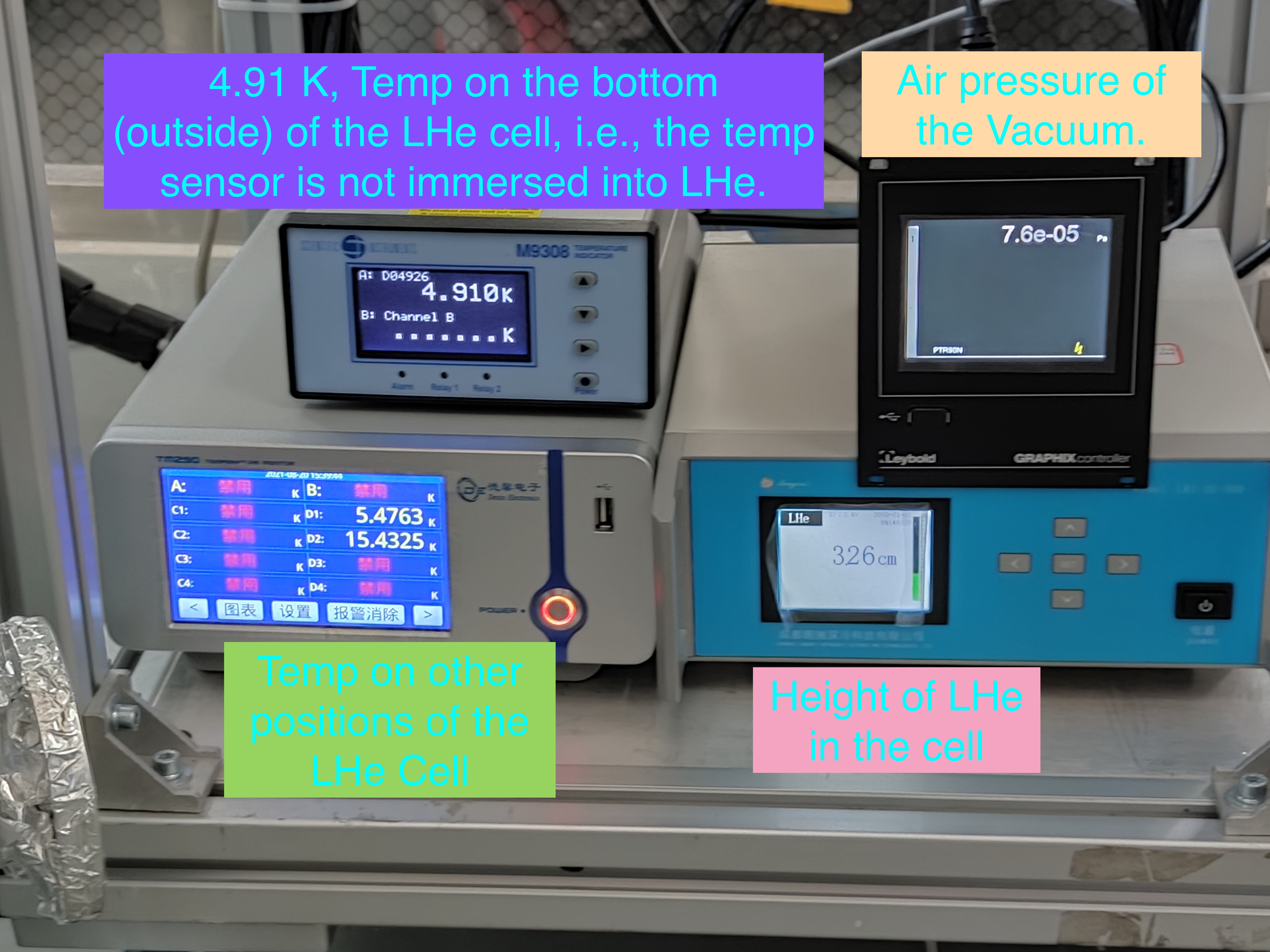}
	\caption{The screen on the upper left shows the temperature on the bottom side of the 30 g detector (The temperature sensor did not immerse into the LHe.), so it is slightly higher than 4.5 K; however, the screen on the lower right clearly shows the height of LHe in the cell is 3.26 cm. The upper right screen is the vacuum inside of the stainless vacuum vessel. The lower left screen shows the temperature on the top side of the 30 g cell (5.4763 K) and the helium gas in the copper vessel (15.4325 K), respectively.}\label{LHeTemperatureAchieved}
\end{figure}

To cool the 30 g detector down to $\sim$4 K, one must first check the vacuum of the whole detector system. If there is a vacuum leaking, it will cause a cryogenic leakage. Consequently, the detector can not be cooled down to the LHe temperature. According to our experiences, there exist two types of vacuum leakage. One is a mechanical failure; this kind of failure could be visible or invisible to the eyes, as mentioned in reference~\cite{ALETHEIA21}. Another is the different Coefficients of Linear Thermal Expansion (CLTE) for closing materials, for instance, the mental anode and the PTFE isolation part surrounding the anode on an HV feedthrough. For the second type of failure, the feature is that the detector works well at RT but will show a leakage as long as the detector is cooled. The reason is that the anode and the isolation are different materials, for instance, copper and PTFE, which have different CLTEs; therefore, their shrinking percentages are not the same at low temperatures; as a result, a tiny gap exists between the anode and the isolation, showing up as a vacuum leakage. After solving variant types of problems~\cite{ALETHEIA21}, we eventually made it: successfully cooled the 30g-V1 detector down to LHe temperature in the summer of 2021, as shown in Fig.~\ref{LHeTemperatureAchieved}.

\begin{figure}[!t]	 
	\centering
        \includegraphics[width=4.0in, angle = 0]{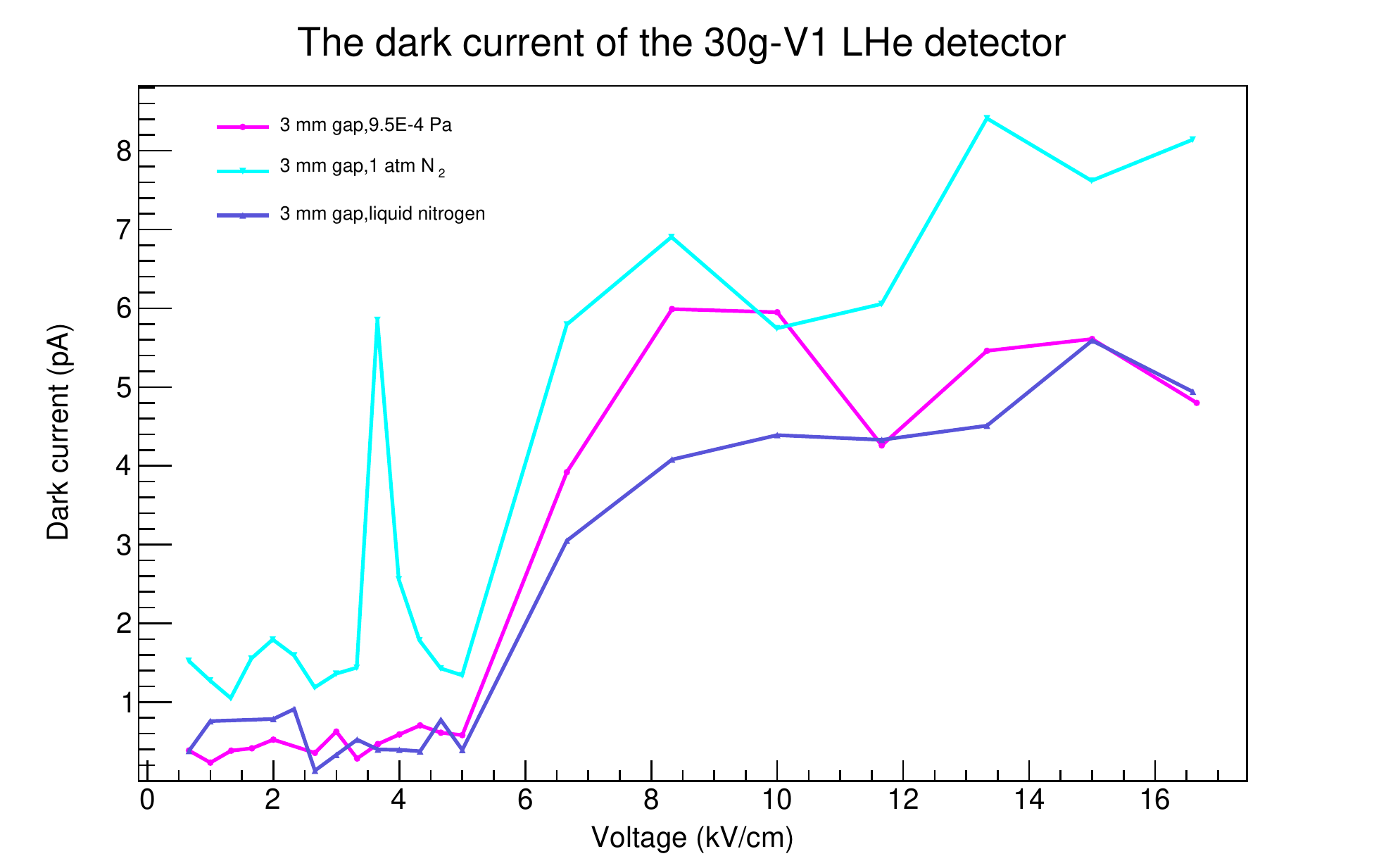}
	\caption{The dark current of the 30g-V1 LHe detector when filled with vacuum, 1 atm N$_2$, and liquid nitrogen, under an HV up to 17 kV/cm.}\label{fig_30gLHeDarkCurrentTest}
\end{figure}

We also built a test benchmark~\cite{ALETHEIA21} to measure the dark current of the 30g-V1 LHe detector under different circumstances: (i) in a vacuum; (ii) filled with 1 atm nitrogen gas; (iii) filled with Liquid Nitrogen (LN). 
The detector's dark current was $< 10 $ pA when the detector was filled with these materials under an HV field up to 17 kV/cm applied. Fig~\ref{fig_30gLHeDarkCurrentTest} shows the results.

\subsection{Testing FBK SiPMs}\label{secProgressALETHEIASubTestingSiPMs}

As mentioned above, SiPMs are critical for the ALETHEIA project. Although variant measurements already demonstrated that certain types of Hamamatsu SiPMs could work around 4 K or even lower~\cite{Cardini14, Iwai19, ZhangSiPM2022}, we have not found any direct report on FBK SiPMs can work at LHe temperatures, which was confirmed by a senior FBK scientists~\cite{privateDiscussionFBKGola}. Therefore, we first tested the I-V curve of more than ten FBK SiPMs with a GM cryocooler to see whether FBK SiPMs can work at 4 K or not; after that, we tested the SiPMs being lighted up with a 450 nm LED at 4 K~\footnote{Calibrating a TPC with LEDs is a conventional method, which has demonstrated by LXe and LAr experiments. However, nobody has ever tested an LHe TPC with LEDs. Therefore, we want to know the viability of such a calibration; as a first step, we would like to know whether a LED can work at 4 K or not.}. Finally, at RT, we tested the SiPMs with a Cs-137 $\gamma$ ray irradiated scintillator, LaBr3:Ce. The SiPM-related tests introduced here are sort of ``qualitative'' measurements in the sense of checking whether FBK SiPMs and the test benchmark's functionality at 4 K~\footnote{All of the FBK SiPM-related tests  in this paper are only shown in our data and do not represent the official results of the FBK company.}. We will perform ``quantitative'' calibrations on SiPMs and the LHe detector in the future.\\

We found the results of FBK SiPM's I-V tests are surprisingly good, though we can not interpret the observations satisfactorily right now.

\begin{figure}[!t]	 
	\centering
        \includegraphics[width=3.2in, angle = 0]{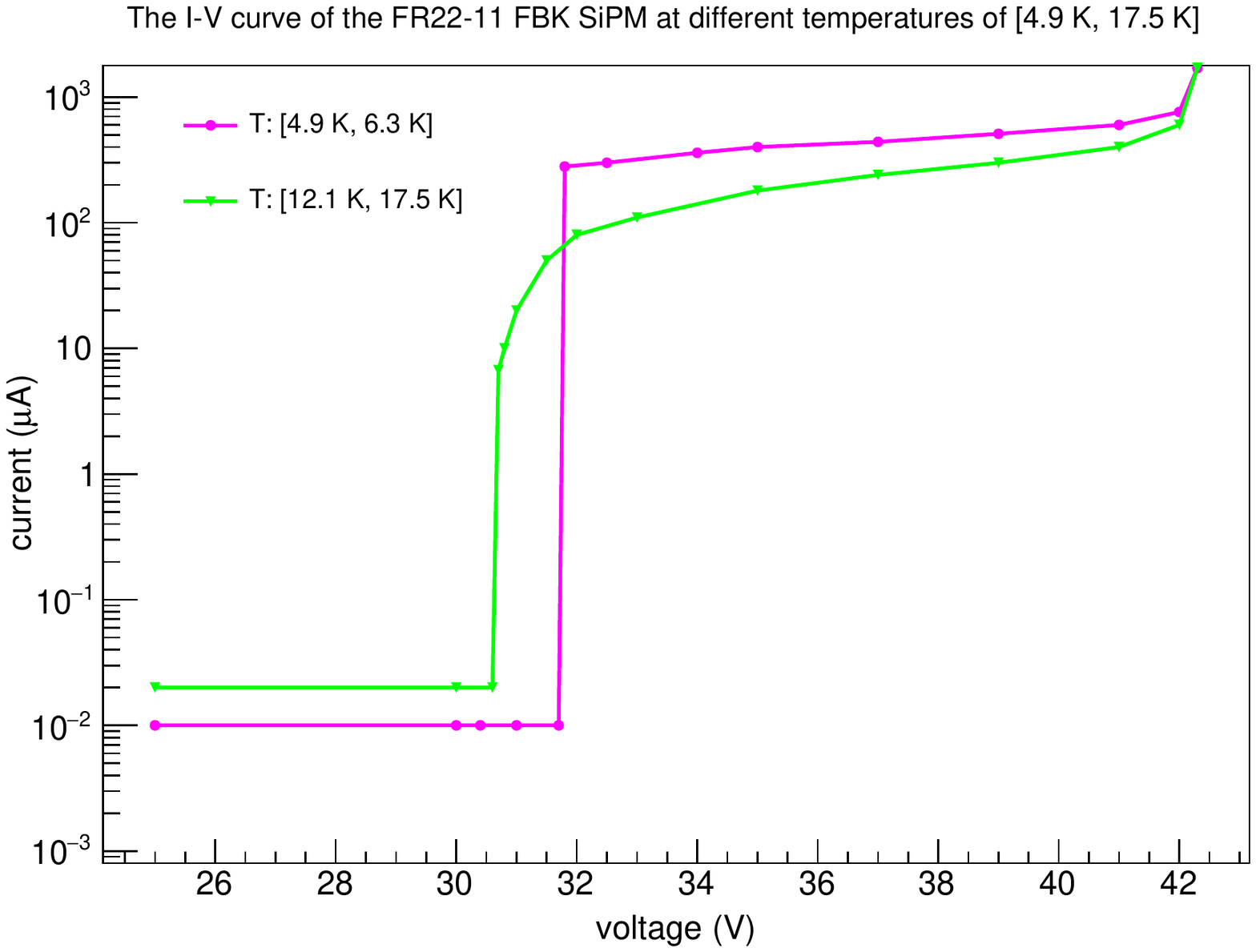}
	\caption{The I-V test for the FR22-11 FBK SiPM at temperatures of [ 4.9 K - 6.3 K] and [12.1 K - 17.5 K]. When the external voltage applied to a SiPM is greater than the breakdown voltage 32 V, there will exist a $\sim$ mA level current, which will increase the temperature of the SiPM under test; when we switch off the voltage, the temperature backs down. That is the reason why for each testing curve, there exists a temperature interval instead of a fixed temperature value.}\label{fig_FBKSiPM11IVTest5To50K}
\end{figure}

\begin{figure}[!t]	 
	\centering
        \includegraphics[width=3.2in, angle = 0]{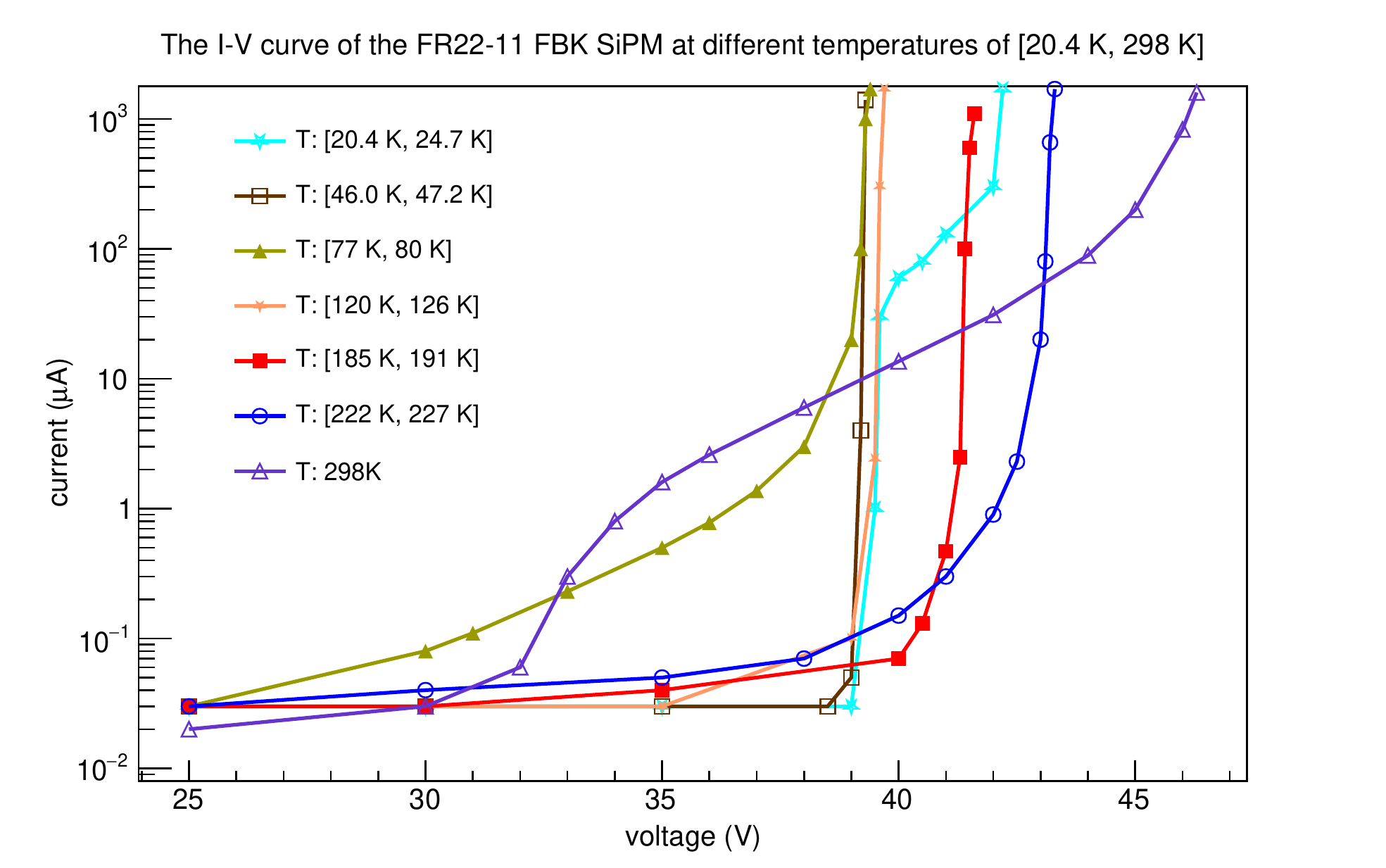}
	\caption{The I-V test for the FR22-11 FBK SiPM at temperatures between [ 77 K, 298 K]. Each testing curve has a temperature interval instead of a fixed temperature. Please refer to Fig.~\ref{fig_FBKSiPM11IVTest5To50K} or main text for the interpretation.}\label{fig_FBKSiPM11IVTest77To298K}
\end{figure}

(a) Roughly a 10 V plateau existed when the SiPMs' temperature was 4 - 20 K. As an example, Fig.~\ref{fig_FBKSiPM11IVTest5To50K} shows such a plateau. However, the plateau did not show up at other higher temperatures, even at RT, as shown in Fig.~\ref{fig_FBKSiPM11IVTest77To298K}.

(b) For a SiPM having the 10 V plateau, the typical breakdown voltage showing at 4 - 20 K is the same as RT. For instance, we can see the breakdown voltage at 4 K and RT are both $\sim$ 32 V, as shown in Fig.~\ref{fig_FBKSiPM11IVTest5To50K} and the ``298 K'' one in Fig.~\ref{fig_FBKSiPM11IVTest77To298K}. 

(c) This welcomed plateau exists in most but not all FBK SiPMs; among the ten SiPMs we tested so far, we found eight of them have such a feature.

(d) Among the eight SiPMs having the 10 V plateau, we randomly selected one and tested it a second time two days later by cooling it down to the same low temperature; the plateau does not change a bit.

Be aware, please, for each curve shown in Fig.~\ref{fig_FBKSiPM11IVTest5To50K} and Fig.~\ref{fig_FBKSiPM11IVTest77To298K}, the temperature is not a fixed value; instead, it is a temperature range. For instance, the temperature corresponds to the magenta curve in Fig.~\ref{fig_FBKSiPM11IVTest5To50K} is between 4.9 K and 6.3 K. This is because during the test, when the applied voltage is greater than the breakdown voltage of 32 V, a $\sim$ mA level current will flow through the SiPM, which will warm up the SiPM under test, therefore, increase the temperature. Conversely, switching off the voltage makes the temperature back down. That is why there exists a temperature fluctuation for each test instead of a fixed value.

\begin{figure}[!t]	 
	\centering
        \includegraphics[width=3.0in, angle = 0]{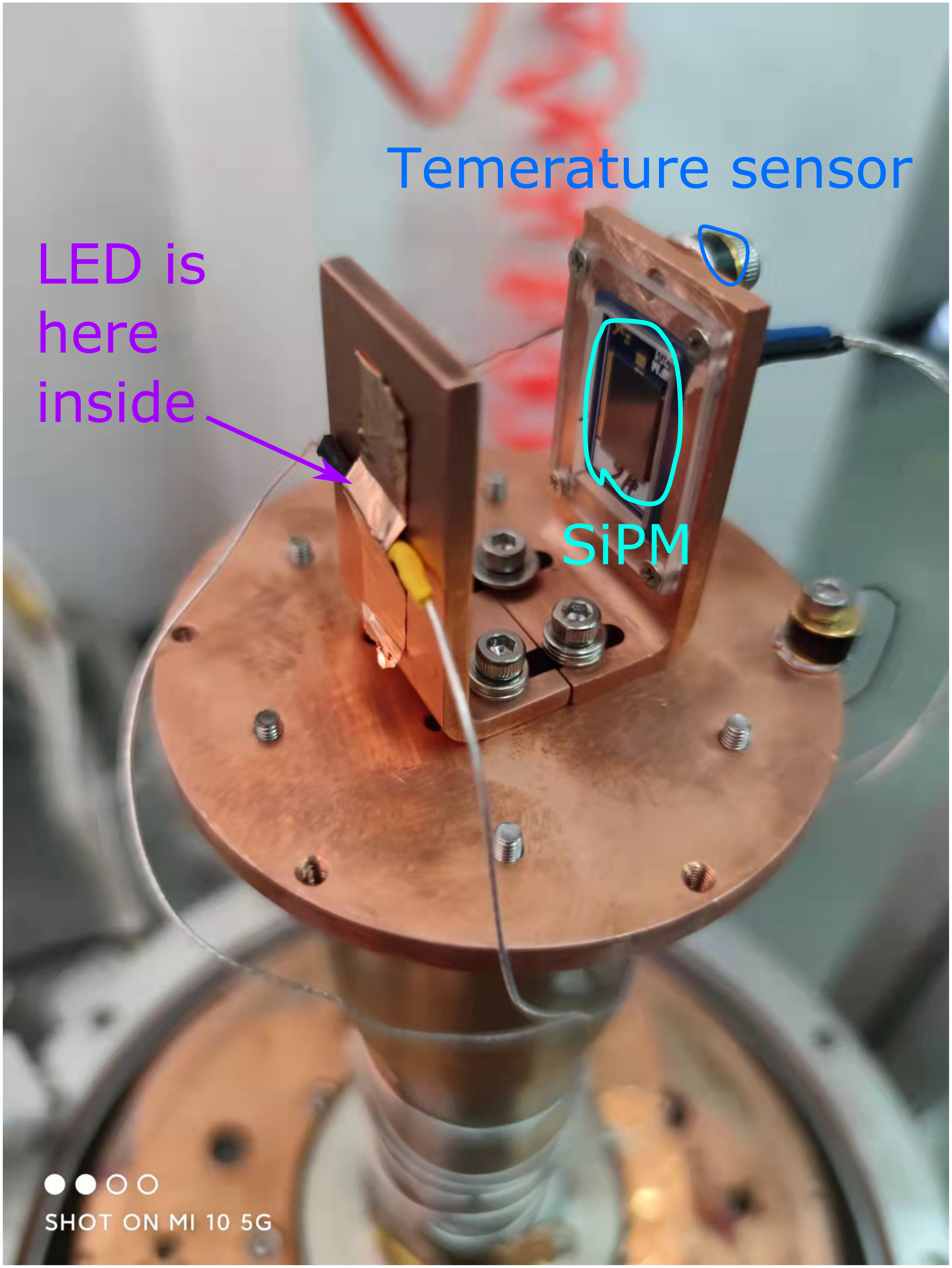}
	\caption{The experimental setup to test SiPMs being lighted up with a 450 nm LED at 4 K.}\label{fig_FBKSiPMLEDTest4KSetup}
\end{figure}

The setup used to test the SiPMs with a 450 nm LED at 4 K is shown in Fig.~\ref{fig_FBKSiPMLEDTest4KSetup}. Actually, for the I-V test mentioned above, we implemented the same test bench; the only difference is that the LED does not light up for the I-V test.

\begin{figure}[!t]	 
	\centering
        \includegraphics[width=3.0in, angle = 0]{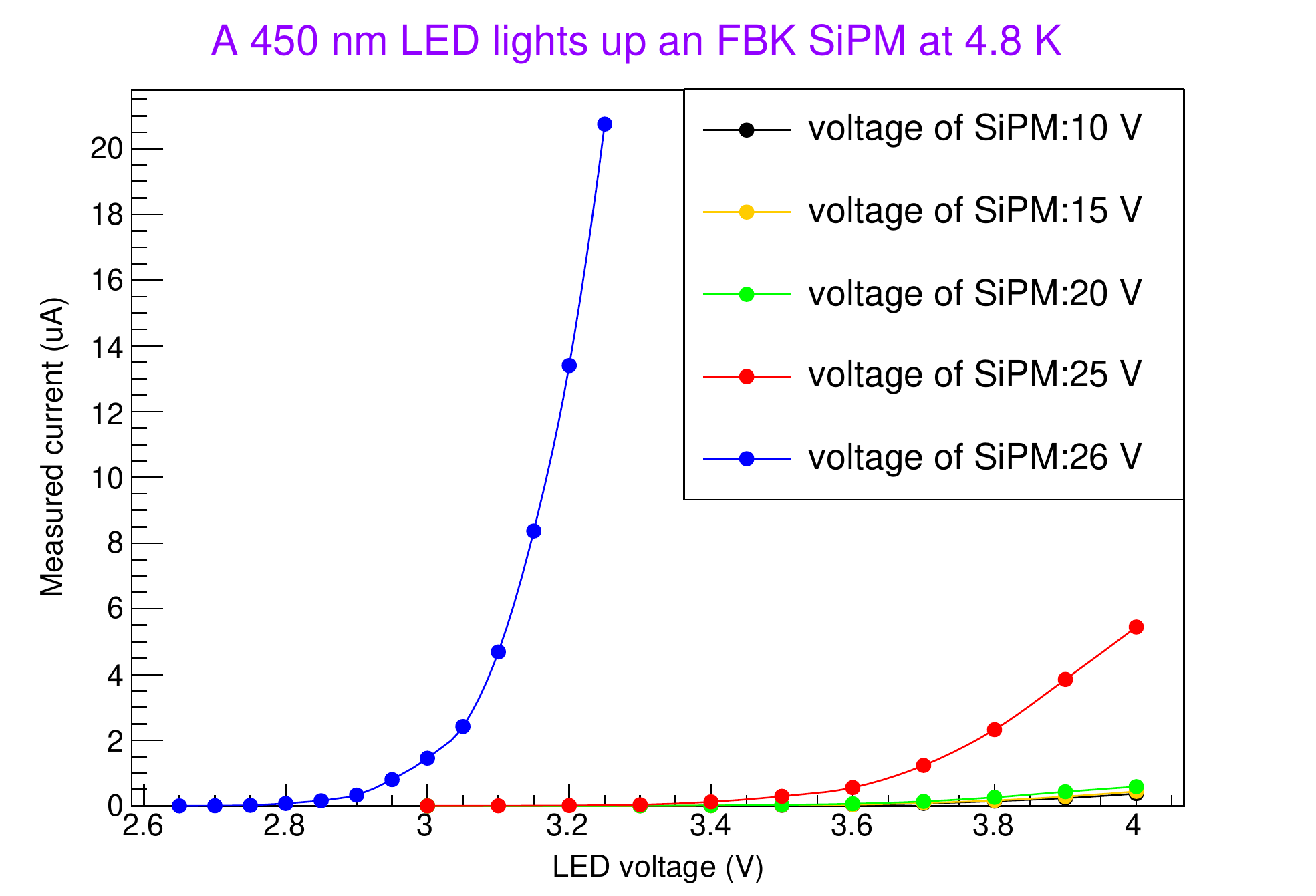}
	\caption{The current changes along with the voltage applied to the SiPM and the LED. Tested at 4.8 K.}\label{fig_FBKSiPMLEDTest4KResults}
\end{figure}

The LED emits 450 nm light. Before testing it at 4 K temperature, we first put the LED into an LN dewar to see whether it would damage or not. We took it out of LN a few hours later and could not find visible damage on the surface. Testing it at RT, the LED can emit light normally as never been immersed into LN before, which gives us the confidence to cool it with the SiPM to LHe temperature. At 4 K, we observed that the measured current changes along with the voltage being applied to the SiPM and the LED, although at the highest voltage under test, 26 V, is still below the breakdown voltage of the SiPM ($\sim$ 32 V), as shown in Fig.~\ref{fig_FBKSiPMLEDTest4KResults}. This test gives us the confidence to use a LED to calibrate an LHe TPC in the future, as those LXe and LAr TPCs have demonstrated.

\begin{figure}[!t]	 
	\centering
        \includegraphics[width=3.0in, angle = 0]{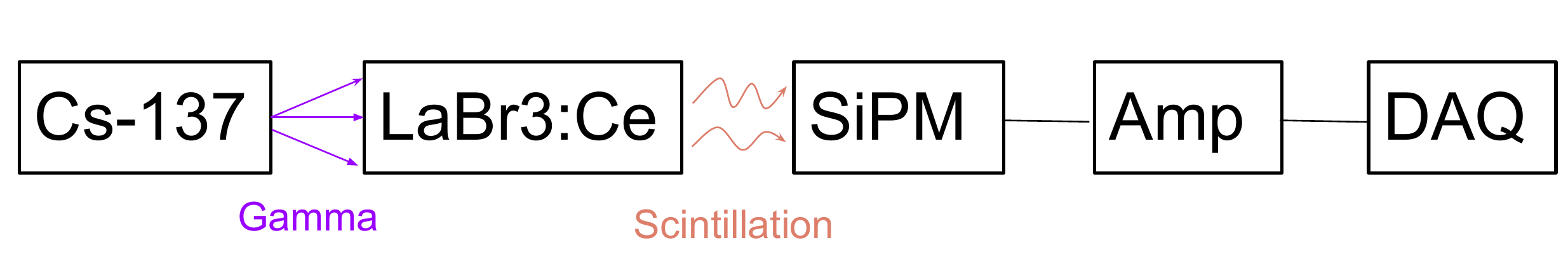}
	\caption{The schematic drawing of a SiPM tested with a Cs-137 $\gamma$ ray irradiated scintillator, LaBr3:Ce. Tested at RT.}\label{fig_FBKSiPMTestWithScintillator}
\end{figure}

At RT, we also tested SiPMs with a Cs-137 $\gamma$ ray irradiated scintillator, LaBr3:Ce. The schematic drawing of the test is shown in Fig~\ref{fig_FBKSiPMTestWithScintillator}. LaBr3:Ce emits scintillation peaked near 400 nm. The Amp is a dedicatedly designed electronic readout, as will be introduced in the following~\ref{secProgressALETHEIASubElectronics}. The DAQ is the Tektronix 70804C Digital Phosphor Oscilloscope, with a bandwidth of up to 8 GHz and a sample rate of 25 GS/s.

\begin{figure}[!t]	 
	\centering
        \includegraphics[width=3.0in, angle = 0]{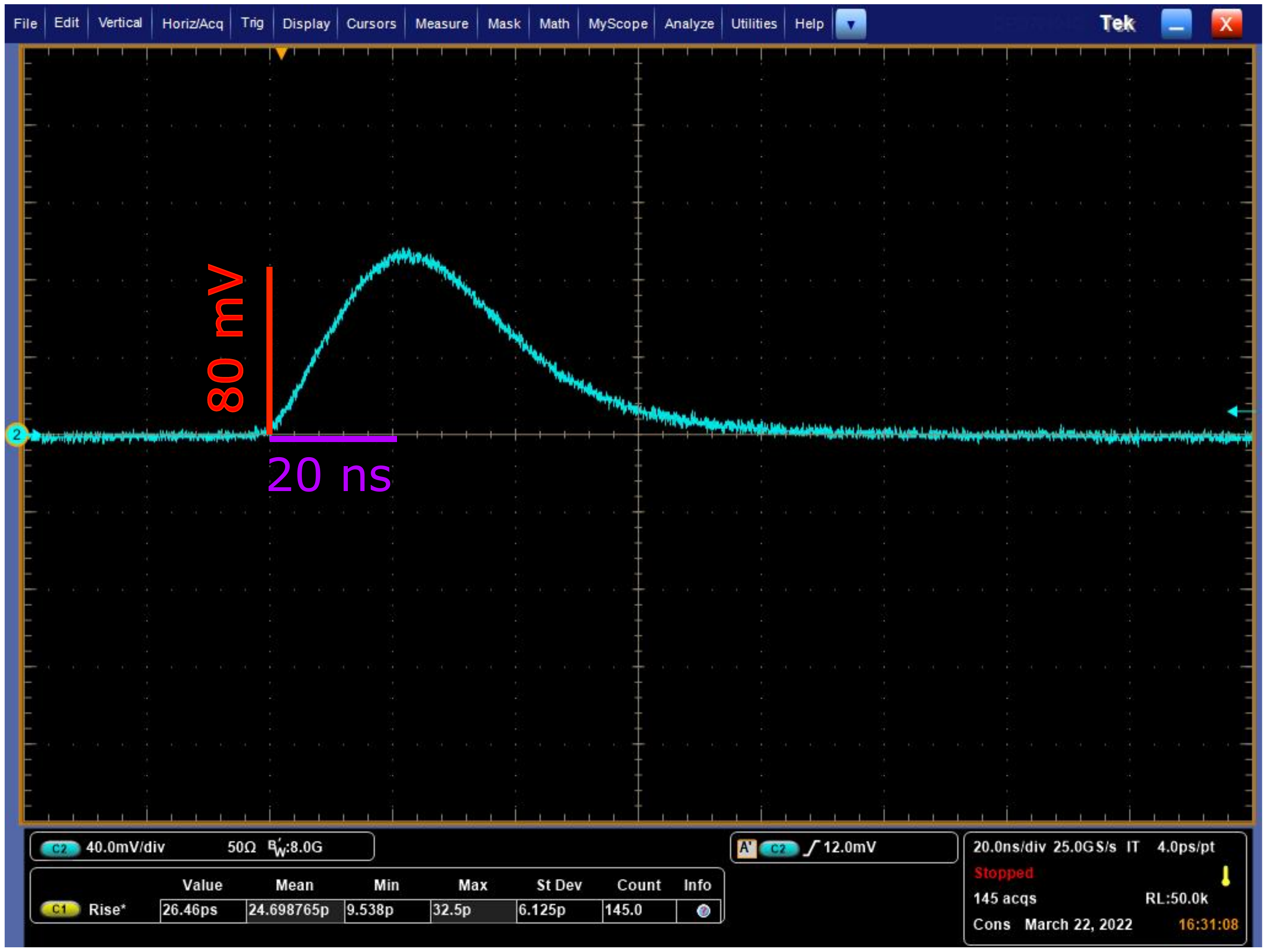}
	\caption{A typical analogy signal observed on the screen of an Oscilloscope. The test setup is shown in Fig.~\ref{fig_FBKSiPMTestWithScintillator}.}\label{fig_FBKSiPMTestTypicalSignals}
\end{figure}

With the setup of Fig.~\ref{fig_FBKSiPMTestWithScintillator}, a typical signal observed on the screen of an oscilloscope is shown in Fig.~\ref{fig_FBKSiPMTestTypicalSignals}.

\subsection{TPB Coating}\label{secProgressALETHEIASubTPBCoating}

As demonstrated in references~\cite{McKinsey03, Ito2012, ItoSeidel13, Seidel14, Ito16, Phan20, SpiceHeRald21}, TPB (1,1,4,4-tetraphenyl-1,3-butadiene) is capable of working at LHe temperature. According to reference~\cite{Benson18}, $\sim 3~ \mu$m is the most appropriate thickness for TPB in terms of maximizing light yield. The TPB layer in DEAP-3600 detector is 3 $\mu$m~\cite{Broerman2017}, though DarkSide-50 and DarkSide-20k have a much more thicker TPB layer up to $\sim 200~ \mu$m TPB~\cite{DarkSide20k17, AgnesDarkSide502015}. 

\begin{figure}[!t]	 
	\centering
        \includegraphics[width=2.0in, angle = 0]{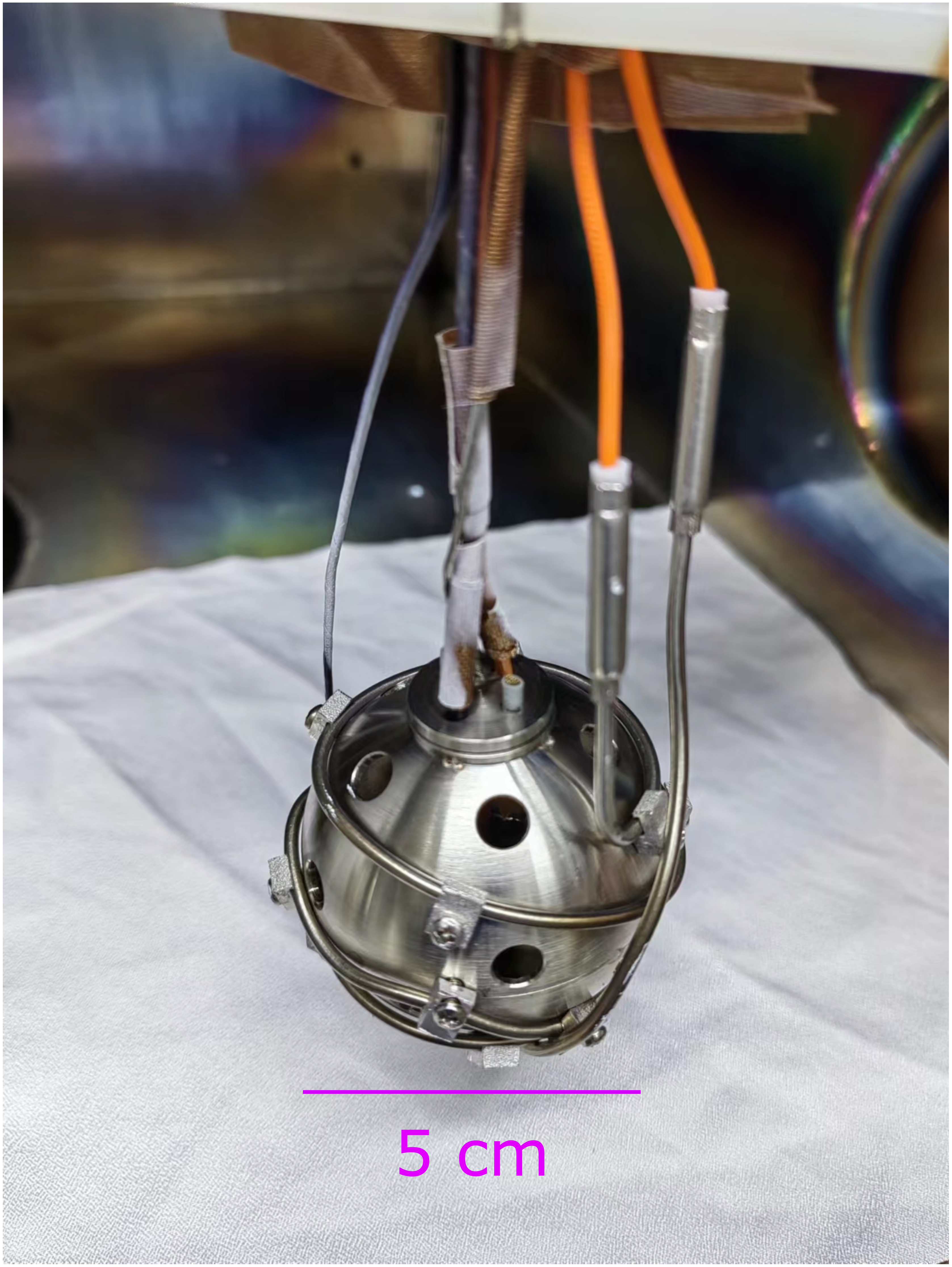}
	\caption{The source sphere used for coating. Inside of the sphere is a crucible which contains TPB powder. There are twenty $\phi ~6$ mm holes evenly distributed on the surface of the $\phi ~5$ cm sphere.}\label{fig_TPBCoatdingSourceSphere}
\end{figure}

We followed the coating process introduced in references ~\cite{PollmannPhDThesis, BroermanMasterThesis, Broerman2017}. However, we cannot mimic the process shown in the papers in our detector directly since (a) our detector shape is cylindrical, not a sphere, and (b) the diameter and height of our detector are only 10 cm, while the DEAP detector's diameter is 1.7-meter. So we must design and build an appropriate coating apparatus to adapt to our detector. Among other technical challenges, the most critical one is building a source. The source here is a small sphere with several holes on the surface; in the center of the sphere is a crucible to contain TPB powder, which will vaporize into the gas after being heated. The tricky of designing such a source is that it must make sure TPB molecules move slowly enough to ensure they scatter each other sufficiently inside the source before randomly finding a hole to escape. In our setup, twenty $\phi ~6$ mm holes are evenly distributed on the surface of the $\phi ~5$ cm sphere, as shown in Fig~\ref{fig_TPBCoatdingSourceSphere}. This way, TPB molecules would come out of the source isotropically and then deposit on the inner surfaces of a cylindrical detector (diameter and height are both 10 cm) with nearly the same thickness.  

We developed three independent methods to measure the thickness of the coated TPB layer.

(i) A real-time monitoring system shows the thickness change during coating.

(ii) Comparing the mass of a few aluminum sample plates inside the detector before and after coating, then figure out the thickness of TPB on the sample plates.

(iii) Calculating the thickness by the consumed TPB mass and the area of the coated region. 

\begin{figure}[!t]	 
	\centering
        \includegraphics[width=2.0in, angle = 270]{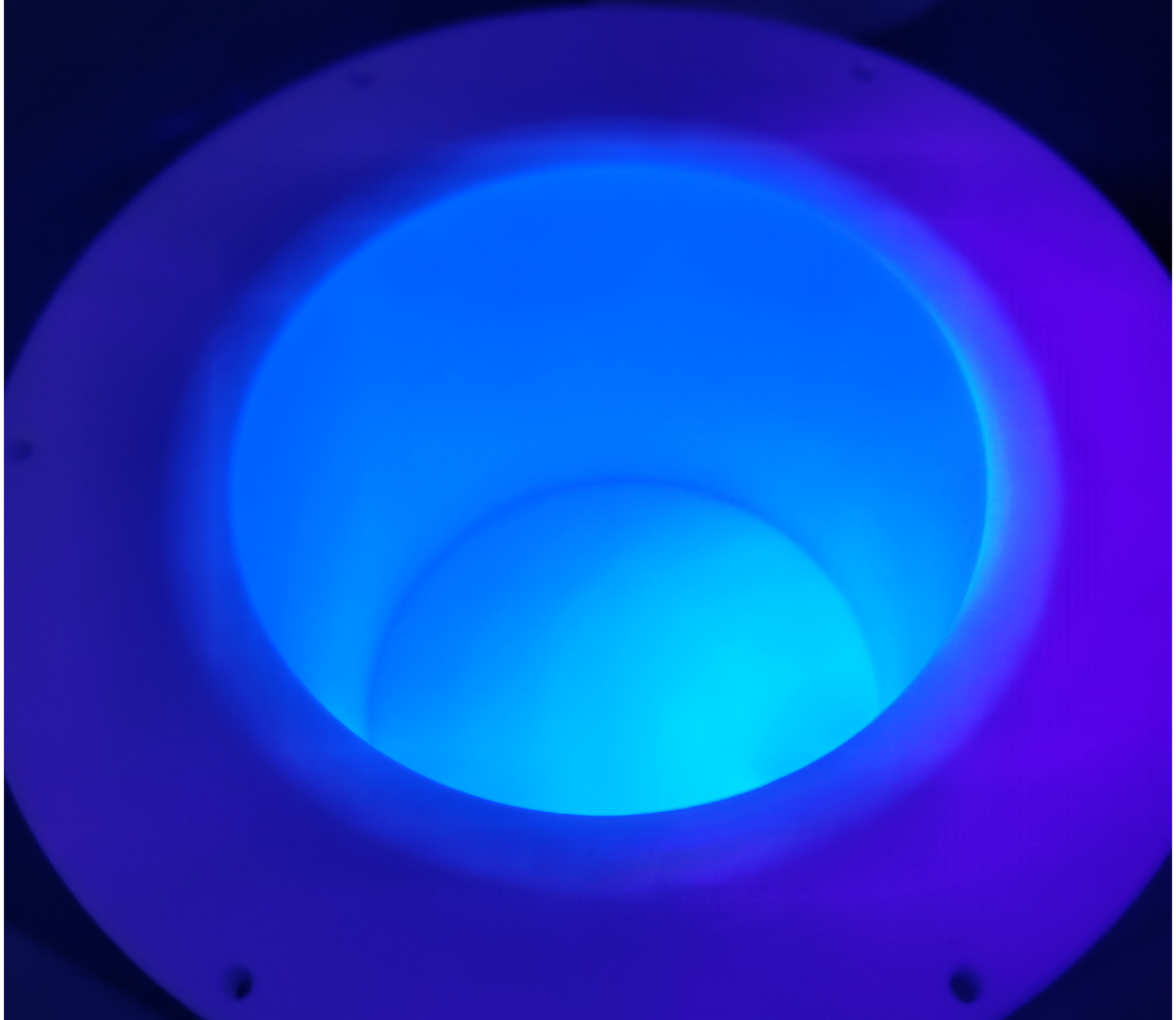}
	\caption{$\sim 3.0~ \mu$m TPB has been coated on the inner wall of a 10-cm PTFE detector (top view).}\label{fig_TPBCoatdingInsideView}
\end{figure}

All of these three methods show consistent thickness results. Fig.~\ref{fig_TPBCoatdingInsideView} shows the 10-cm PTFE detector coated with $\sim 3.0 ~\mu$m TPB. By changing the TPB mass in the crucible, we can, in principle, coat any thickness we need. For details of the TPB coating work, please refer to our papers~\cite{TPBCoutingALETHEIA22}.

In addition, to check whether the TPB coating layers will be damaged at low temperatures, we scanned the surfaces of TPB-coated sample films with an SEM (Scanning Electron Microscope) machine. Some of these sample films had gone through three hours at 4 K temperature in a G-M cryocooler, some were immersed into an LN dewar for 40 hours, and some had no cryogenic exposures. The SEM images do not show any noticeable difference for those films, which gives us extra confidence to implement TPB as a wavelength shifter into our LHe detector. For more details, please refer to our published paper~\cite{TPBCoutingJINSTNov22}.

\subsection{Readout Electronics}\label{secProgressALETHEIASubElectronics}

\begin{figure}[!t]	 
	\centering
        \includegraphics[width=4.0in, angle = 0]{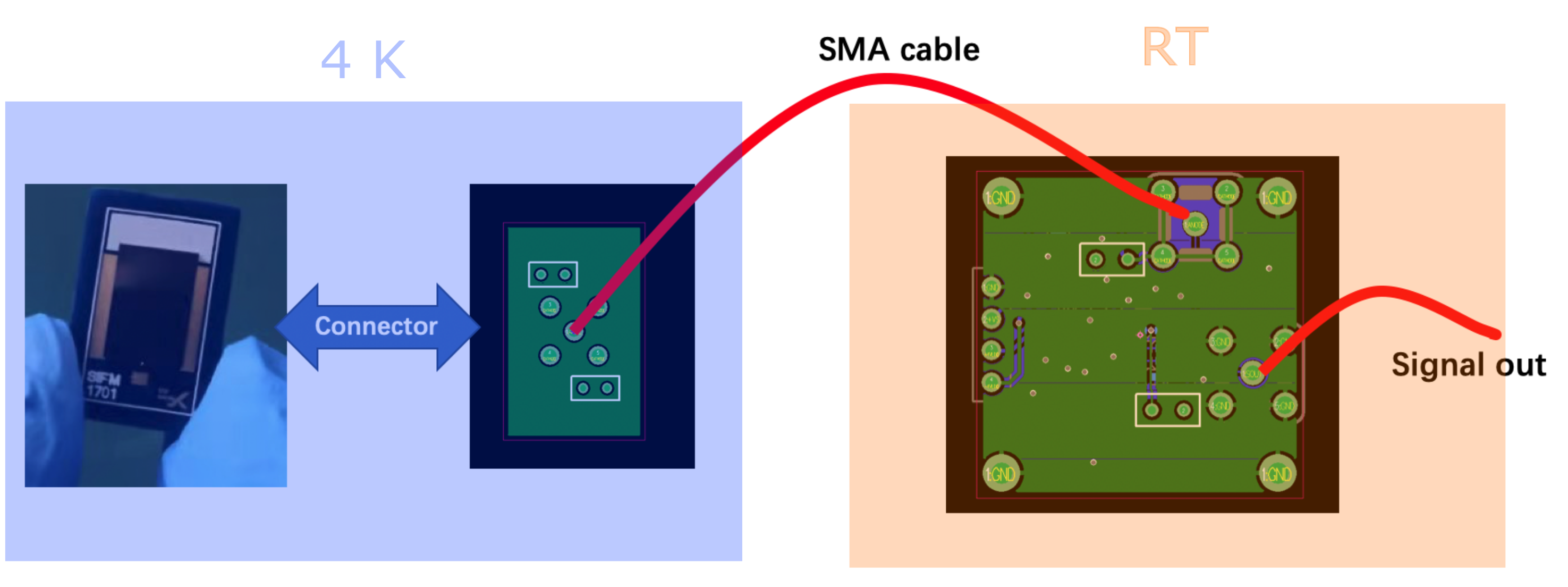}
	\caption{A schematic drawing shows how a SiPM, a connection board, and a pre-amplifier board connect. The SiPM and the connection board will work at 4 K, while the pre-amplifier works at RT.}\label{fig_SiPMElectronicsSchematic}
\end{figure}

We have cooperated with an electronic company to design electronics suitable for FBK SiPMs and our LHe detector. Considering it would be a risk for Integrated Circuits (ICs) to work at the LHe temperature, the pre-amplifier should not work at the temperature. Since the SiPMs must locate very close to LHe, the SiPM's temperature would be around 4 K; to read the SiPMs' current out, we designed a connection board without ICs on it, which can work at 4 K. The pre-amplifier board works at RT with ICs. The two boards are connected via a 1-meter SMA cable. A schematic drawing of such a connection is shown in Fig~\ref{fig_SiPMElectronicsSchematic}. 

\begin{figure}[!t]	 
	\centering
        \includegraphics[width=4.0in, angle = 0]{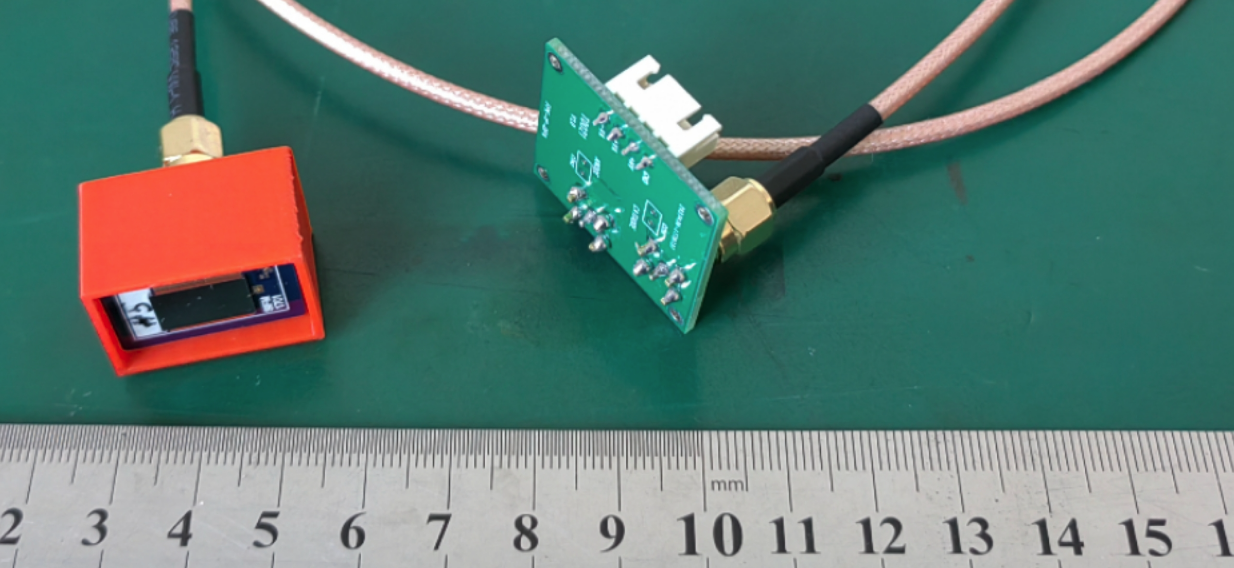}
	\caption{The orange box contains a SiPM and a connection board (the board is invisible in the picture since it is behind the SiPM); a pre-amplifier board connects the connection board with an SMA cable. The orange box will work at LHe temperature, while the pre-amplifier will work at RT.}\label{fig_SiPMElectronicsReal}
\end{figure}

Fig~\ref{fig_SiPMElectronicsReal}  shows the real boards we developed. The connection board is invisible in the plot since it locates behind the SiPM in the orange box. The two boards performed well at RT, as mentioned in section~\ref{secProgressALETHEIASubTestingSiPMs}. We will test them at 4 K in the coming months. 

\section{Summary}\label{secSummaryALETHEIA}

DM is one of the most pressing questions to be understood and answered in fundamental physics today. High-mass WIMPs detection has achieved the cross-section of 10$^{-48} $ cm $^2$, while still no signal has been convincingly observed yet. However, the upper limits on low-mass WIMPs are only $\sim$ 10$^{-38} $ cm $^2$, roughly 10 orders behind high-mass WIMPs searches. The community should shed more light on the low-mass DM than ever.

Filling with the arguably cleanest LHe into the arguably most competitive DM detector, TPC, ALETHEIA is supposed to achieve extremely low backgrounds. Projecting with the LZ detector, for a 1-ton LHe TPC running with 3 years exposure, the ER and NR backgrounds are supposed to be $\sim$ 11 and $\sim$ 0.5, respectively. Such a low background is very helpful in answering the most challenging question in fundamental physics today: the nature of DM, especially under the circumstance that we know nothing about the particle physics character of DM: it could be a NR, or ER, or some other forms. A 3 ton*yr exposure could achieve a cross-section of $\sim$ 10$^{-46} $ cm $^2$, roughly one order below the $^8$B neutrino floor. A panel of world-leading physicists in the field thankfully reviewed the project with very positive comments and comprehensive suggestions in Oct 2019. Following the review panel, we have built a couple of 30 g LHe detectors to verify some essential performances of LHe; we will build a kg-size detector to thoroughly verify the technological routine of LHe TPC in the future.

Since launching the project in the summer of 2020, we have made significant progress in the past two years.

(a) We built a 30 g detector and successfully cooled it down to $\sim$4 K by filling liquid helium. 

(b) We demonstrated that FBK SiPMs are capable of working at $\sim$4 K; and even more surprisingly, we found there exists an up to 10 V plateau above the breakdown voltage for temperatures among $\sim$ 4 - 20 K; i.e., if the breakdown voltage is 32 V, the plateau would be 32 - 42 V, while such a plateau does not exist for temperatures higher than $\sim$ 20 K until RT. We observed eight SiPMs that have the plateau among the ten tested. We currently are not capable of fully understanding the observations.

(c) We developed a coating process that enables us to coat $\sim 3 ~ \mu$m TPB evenly on the inner walls of a 10-cm cylindrical detector; We are confident to coat TPB on larger detectors with the variant thickness on need. 

(d) We designed and developed a dedicated electronics system to read out FBK SiPMs; We integrated it into the DAQ system and took data successfully.

(e) We built a few benchmarks to cope with multiple tests: I-V tests for SiPMs both at RT and 4K, data taking with an oscilloscope for SiPMs being lighted with a LED, or the scintillation coming from a scintillator irradiated by a $\gamma$ source.

(f) The 30g-V2 detector is currently under assembly.

In summary, after working on the project for a couple of years, we are more confident to build an LHe TPC successfully, although we understand that much R\&D work is ahead.


\bmhead{Acknowledgments}

We thank the professors who flew to Beijing in Oct 2019 to participate in the DM workshop and reviewed the project: Prof. Rick Gaitskell, Prof. Dan Hooper, Dr. Takeyaso Ito, Prof. Jia Liu, Prof. Dan McKinsey, and Prof. George Seidel. The panel also thankfully addressed major R\&D questions and recommendations. This paper referred to the workshop talks and the review document. 

Junhui Liao thanks Prof. Weiping Liu for helping settle down at CIAE in the context of the unexpected COVID pandemic. Junhui Liao would also thank the support of the ``Yuanzhang'' funding of CIAE to launch the ALETHEIA program.

\bmhead{Data Availability Statement}
The datasets generated during and/or analysed during the current study are not publicly available due to a preliminary research stage but are available from the corresponding author on reasonable request.

\bibliography{1ALETHEIA-RV1}

\end{document}